\documentclass[letterpaper]{article}

\usepackage{fullpage,times}

\usepackage{times}  
\usepackage{hyperref}  
\usepackage{graphicx} 
\usepackage{natbib}  
\usepackage{caption} 
\usepackage{algorithm}
\usepackage{newfloat}
\usepackage{listings}

\usepackage{latexsym}
\usepackage[T1]{fontenc}
\usepackage[utf8]{inputenc}
\usepackage{microtype}
\usepackage{inconsolata}
\usepackage{lineno}
\usepackage{algpseudocode}
\usepackage{mathtools}
\usepackage{amsthm}
\usepackage{amssymb}

\usepackage{multirow}
\usepackage{multicol}
\usepackage{color, colortbl}
\usepackage{tcolorbox}
\usepackage{xcolor}
\usepackage{tabularx}
\usepackage{amsmath}
\usepackage{wrapfig}
\usepackage{enumitem}
\usepackage{etoolbox}
\usepackage{subcaption}
\usepackage{hhline}
\usepackage{booktabs} 
 \usepackage{arydshln}
 \usepackage{siunitx}
 \usepackage{xcolor}

\definecolor{Gray}{gray}{0.85}
\definecolor{LightGray}{gray}{0.90}
\definecolor{LLightGray}{gray}{0.95}
\definecolor{Red1}{RGB}{235,110,110}
\definecolor{Red2}{RGB}{247,192,192}

\DeclareRobustCommand{\papername}{Tacit-TTS}

\title{\papername: From Autoregressive Decoding to Masked Prediction for Efficient Transcript-Free Voice Cloning}

\author{
  Jian Chen\\
  Dolby Laboratories\\
  \texttt{Jian.Chen@dolby.com}
  \and
  You Zhang\\
  Dolby Laboratories\\
  \texttt{Neil.Zhang@dolby.com}
  \and
  Mark Vinton\\
  Dolby Laboratories\\
  \texttt{Mark.Vinton@dolby.com}
}

\date{}

\IfFileExists{main_backup.aux}{
  \message{We saw a default backup.aux file, let's use it instead of the main aux file.}
  \nofiles 
  \makeatletter
  \input{main_backup.aux}
  \makeatother
}{}

\begin{document}

\maketitle

\begin{abstract}
TTS systems with autoregressive semantic modeling have demonstrated strong zero-shot voice cloning performance and rich expressive variation, but their sequential decoding incurs substantial latency. Non-autoregressive alternatives offer much faster generation, yet often rely on more restrictive reference conditioning, such as requiring transcripts of the reference speech during inference. We present \papername{}, an efficient transcript-free zero-shot voice cloning system distilled from IndexTTS2. \papername{} replaces autoregressive text-to-semantic decoding with masked non-autoregressive generation, introduces training-free acoustic length estimation, and accelerates the flow-matching renderer through ReFlow distillation. Across two English and two Mandarin datasets, \papername{} achieves competitive zero-shot quality while generating speech over 10$\times$ faster than IndexTTS2 for utterances longer than 5 seconds. Its transcript-free conditioning further supports cross-lingual and non-lexical references. We validate this capability using references from eight other languages, infant babble, and synthetic gibberish, where transcript-dependent systems often degrade or fail due to unreliable ASR transcripts.
\end{abstract}
\section{Introduction} \label{sec:introduction}

\begin{wrapfigure}{r}{0.5\linewidth}
    \centering
    \vspace{-1.2em}
    \resizebox{\linewidth}{!}{\includegraphics{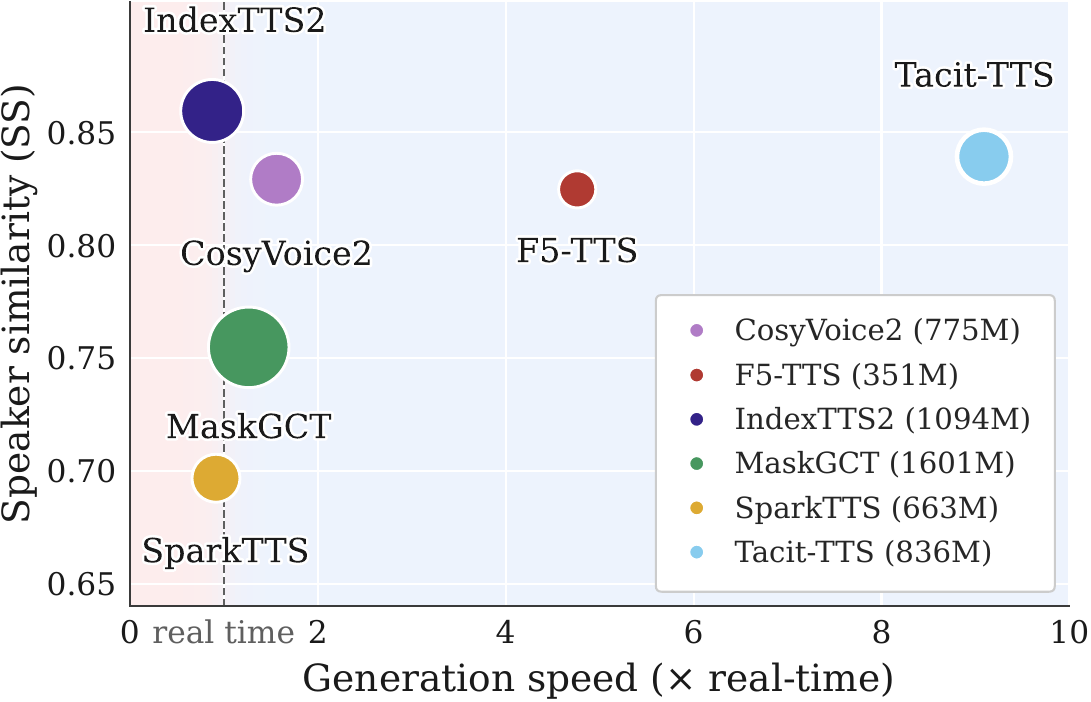}}
    \caption{Efficiency–speaker similarity trade-off of \papername{} and baselines. Marker size indicates the number of generation-module parameters.}
    \vspace{-1em}
    \label{fig:eff-ss-intro}
\end{wrapfigure}

As generative AI increasingly interacts with users through speech, the ability to generate a convincing and personalized voice has become an important part of human-AI interaction. Modern voice cloning systems can reproduce the identity and expressive characteristics of an unseen speaker from only a short reference recording, enabling applications such as multilingual speech translation and dubbing~\citep{barrault2023seamless}, personalized virtual agents~\citep{casanova2022yourtts}, and healthcare and assistive applications~\citep{jreige2009vocalid}. Beyond intelligibility, preserving speaker identity and emotion makes synthesized speech more natural and engaging~\citep{ju2024naturalspeech, zhou2026emotional}, bringing AI-generated voices closer to human communication.

Achieving this level of fidelity, however, remains expensive in both training and inference. High-quality speaker reproduction and expressive control benefit from large and diverse speech corpora covering many speakers, linguistic content, and speaking styles. Industrial-scale systems such as IndexTTS2~\citep{zhou2026indextts2}, trained on tens of thousands of hours of speech, demonstrate strong zero-shot quality but rely on autoregressive (AR) semantic generation, whose sequential decoding introduces substantial latency. Non-autoregressive (NAR) systems alleviate this bottleneck through parallel generation~\citep{wang2025maskgct,chen2025f5,eskimez2024e2}, but typically require large-scale training and often depend on transcripts of the reference speech. Transcript dependence limits voice cloning when the reference language is unsupported by automatic speech recognition~(ASR) or when meaningful lexical content is unavailable. Fixed-length NAR decoding introduces a separate constraint, requiring the target duration to be determined before generation. These limitations leave a practical gap between high-quality voice cloning and voice cloning that is efficient, transcript-free, and free of upfront duration constraints.

To address these challenges, we propose \papername{}, an efficient transcript-free zero-shot voice cloning system distilled from IndexTTS2. We replace the teacher's autoregressive text-to-semantic (T2S) model with a smaller masked non-autoregressive generator trained on teacher-synthesized data, while retaining its pretrained speaker and emotion conditioning. To enable transcript-free NAR decoding, we introduce a training-free length-control strategy that estimates speaking pace directly from the reference audio and combines it with the syllable count of the target text. We further recover the teacher's continuous latent representation and accelerate the downstream flow-matching S2Mel renderer through ReFlow distillation.

We evaluate \papername{} on two English and two Mandarin datasets, together with perceptual-quality, duration-fidelity, and efficiency analyses. \papername{} achieves competitive zero-shot quality and the highest speaker similarity among the evaluated non-teacher baselines on both English datasets, while maintaining perceptual quality close to IndexTTS2 and generating durations that correlate closely with ground truth. It also generates speech over 10$\times$ faster than IndexTTS2 for utterances longer than 5 seconds. We further evaluate transcript-free conditioning with cross-lingual references from eight languages and non-lexical references including infant babble and synthetic gibberish. In these settings, transcript-dependent systems often degrade or fail because of unreliable ASR transcripts, while \papername{} remains effective across all references.

Our main contributions are summarized as follows:
\begin{itemize}
    \item We replace the autoregressive T2S stage of IndexTTS2 with masked non-autoregressive generation, accelerating the T2S stage by 26.5$\times$, and further accelerate the downstream flow-matching renderer through ReFlow distillation.
    
    \item We enable transcript-free NAR generation through training-free acoustic length estimation and distillation of both discrete semantic codes and continuous latent representations, without access to the teacher's original large-scale training corpus.
    
    \item \papername{} achieves competitive zero-shot quality while supporting cross-lingual and non-lexical references where transcript-dependent systems often degrade or fail.
\end{itemize}



\section{Related Work} \label{sec:related_work}

\paragraph{Autoregressive and masked generation.}
Zero-shot voice cloning synthesizes speech for unseen speakers from a short reference recording. YourTTS~\citep{casanova2022yourtts}, built on the end-to-end VITS backbone~\citep{kim2021conditional}, clones voices from a speaker embedding alone. Most later systems instead adopt a two-stage generation scheme, first mapping text to semantic tokens and then semantic tokens to audio. VALL-E~\citep{wang2023neural} generates codec tokens with an autoregressive language model conditioned on phoneme and acoustic prompt tokens. Seed-TTS~\citep{anastassiou2024seed} scales this autoregressive scheme, and CosyVoice~\citep{du2024cosyvoicescalablemultilingualzeroshot} and Spark-TTS~\citep{wang2025spark} decode semantic tokens with language models. IndexTTS2~\citep{zhou2026indextts2,deng2025indextts}, our distillation teacher, conditions its autoregressive model on w2v-BERT features compressed by a Perceiver into speaker and emotion embeddings. On the non-autoregressive side, masked prediction descends from BERT-style bidirectional language modeling~\citep{devlin2019bert}: MaskGIT~\citep{chang2022maskgit} generates tokens in parallel by iteratively unmasking confident predictions, a scheme formalized as discrete diffusion~\citep{austin2021structured}. MaskGCT~\citep{wang2025maskgct} applies this decoding scheme to the text-to-semantic stage of zero-shot TTS. Our model adopts MaskGCT's masked architecture. However, instead of training the text-to-semantic model on text-audio pairs, we distill it from an autoregressive teacher, and instead of phone-proportional scaling or learned duration predictors~\citep{ren2019fastspeech,ren2020fastspeech}, we estimate the target length from a learning-free acoustic pace~\citep{de2009praat}.

\vspace{-0.5em}
\paragraph{Audio generation.}
The semantic-to-audio stage turns the semantic representation into audio: most systems synthesize acoustic features such as mel spectrograms and apply a vocoder, while others predict discrete codec tokens for a codec decoder. Flow matching~\citep{lipman2022flow} and rectified flow~\citep{liu2022flow} learn velocity fields over straight paths. Voicebox~\citep{le2023voicebox} generates acoustic features with flow matching, and NaturalSpeech~2~\citep{shen2024naturalspeech} pursues latent diffusion in a codec feature space. E2 TTS and F5-TTS~\citep{eskimez2024e2,chen2025f5} forgo the two-stage split and synthesize mel spectrograms directly from text, without an intermediate semantic representation. SoundStorm~\citep{borsos2023soundstorm} and MaskGCT instead generate discrete codec tokens. IndexTTS2's renderer is a flow-matching model requiring roughly 25 Euler steps. Instead of generic knowledge distillation~\citep{hinton2015distilling}, we fine-tune the renderer on the outputs of our text-to-semantic model and apply reflow~\citep{liu2022flow}, integrating the straightened trajectories in 4 to 8 Euler steps. Together with the masked decoder, this yields the efficiency numbers in Section~\ref{sec:eff}.

\vspace{-0.5em}
\paragraph{Transcript dependence.}
Most zero-shot systems, including the non-autoregressive ones, depend on the reference transcript. F5-TTS transcribes the reference with an internal Whisper model. E2 TTS concatenates the reference text into its conditioning. Voicebox requires the prompt's phoneme sequence. MaskGCT and CosyVoice~2~\citep{du2024cosyvoice2scalablestreaming} require the prompt text directly. When a reliable transcript is unavailable, for unsupported languages, non-lexical vocalizations, or speech impairments~\citep{hartman2017devising}, these systems degrade at their input stage, conditioning on empty or hallucinated text. Conditioning on self-supervised representations instead, w2v-BERT~\citep{chung2021w2v} features compressed by a Perceiver~\citep{jaegle2021perceiver} as in IndexTTS2, removes this dependence. Our evaluations with cross-lingual and non-lexical references in Section~\ref{sec:crossling} and~\ref{sec:nonlex} show that transcript-free conditioning remains effective where transcript-dependent baselines degrade or fail.
\section{Method} \label{sec:method}
\begin{figure}[!h]
\centering
\includegraphics[width=\linewidth]{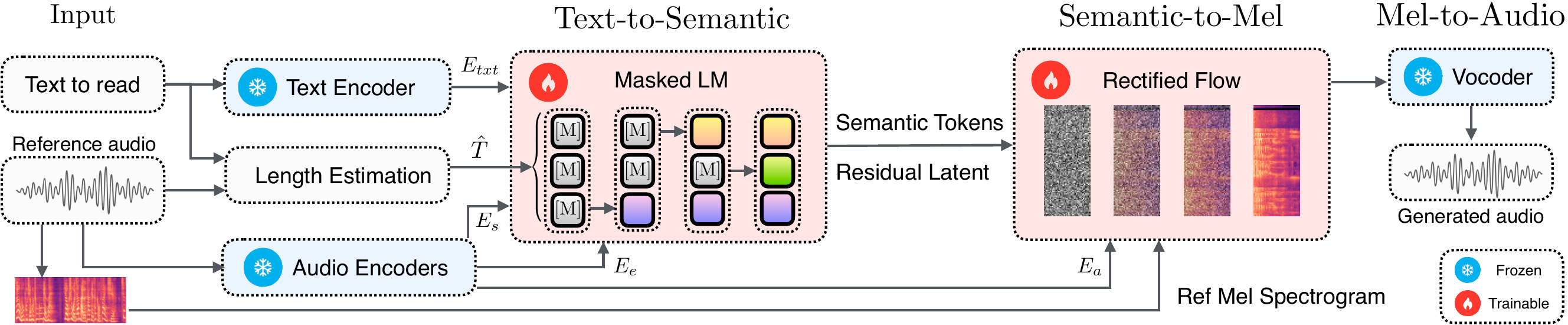}
\caption{Overview of the \papername{} pipeline. Given a reference audio and target text, the encoders extract the conditioning representations, the text-to-semantic model generates semantic tokens via masked-LM generation, and the semantic-to-mel model converts them into a mel spectrogram for waveform synthesis by the vocoder.}
\label{fig:overview}
\end{figure}

\papername{} follows the two-stage generation paradigm adopted by recent zero-shot voice cloning systems, as illustrated in Figure~\ref{fig:overview}. A text to semantic (T2S) module first predicts a sequence of semantic tokens from the reference speech and target text, followed by a semantic to audio (S2A) module that synthesizes the final speech waveform. In the T2S stage, we retain the disentangled conditioning mechanism of IndexTTS2 for speaker identity and emotion control while replacing the autoregressive (AR) semantic decoder with a non-autoregressive (NAR) Transformer. Unlike existing transcript-based NAR systems such as MaskGCT, our model does not require transcripts of the reference speech, eliminating the additional computational overhead of ASR transcription. In the S2A stage, we adopt the flow-matching speech renderer from IndexTTS2 and further accelerate it through reflow distillation, improving inference efficiency while preserving synthesis quality.

\subsection{Text-to-Semantic Generation}
\label{sec:t2s}

The proposed T2S model is implemented as a bidirectional DiT conditioned on masked semantic tokens, text embeddings, and reference-speech embeddings extracted by pretrained encoders (Appendix~\ref{app:percievers}). The reference conditioning consists of a fixed-length speaker embedding sequence $E_s\in\mathbb{R}^{32\times d}$ and an emotion embedding $E_e\in\mathbb{R}^{d}$, which are prepended as prefix tokens to the Transformer. The target text is embedded as $E_{\mathrm{txt}}$, upsampled to the target semantic length $T$, and fused with the semantic token embeddings through concatenation followed by a linear projection. This keeps the Transformer sequence length at $T$ rather than $T+N_{\mathrm{text}}$ and provides an explicit text-to-semantic alignment prior, following the length-regulation principle used in NAR TTS~\citep{ren2019fastspeech, ren2020fastspeech}. Specifically, starting from a fully masked semantic sequence $\mathbf{y}^{0}$ ($\mathbf{y}^{0}_{i}=\textsc{mask}, \forall i$), the model predicts semantic tokens over a fixed number of decoding iterations. At iteration $m$, it computes
$\hat{\mathbf{y}}^{m}=f_{\text{t2s}}\!\left(\mathbf{y}^{m-1},E_{\mathrm{txt}},E_s,E_e\right)$
for all masked positions in parallel. After each iteration, the most confident predictions are committed and the remaining masked positions are refined until all tokens are determined. The number of forward passes is therefore fixed and independent of output length, unlike AR decoding, which scales linearly with semantic sequence length.

\subsubsection{Training-Free Length Control} \label{sec:pace}
Unlike AR T2S models that generate until a stopping token, our NAR T2S model requires the target semantic length $T$ before decoding. In zero-shot synthesis, $T$ depends on both the amount of linguistic content in the target text and the speaking pace of the reference speech. Existing length estimation strategies typically rely on the reference transcript or a learned duration model. To preserve transcript-free inference, we instead estimate content from the target text and the speaking pace directly from the reference waveform.

Specifically, let $D_r$ denote the reference duration after trimming leading and trailing silence while preserving internal pauses, and let $f$ be the semantic codec frame rate, giving $T_r=D_r f$ semantic positions. Let $N_r^{a}$ and $N_x^{t}$ denote the syllable counts estimated from the reference audio and target text, respectively. We define the reference pace factor as $\rho_r=T_r/(\kappa N_r^{a})$. To prevent excessively long or short generations caused by noisy syllable estimates or atypical reference speech, we constrain it as $\tilde{\rho}_r=\mathrm{clip}(\rho_r,\rho_{\min},\rho_{\max})$. The target semantic length is then estimated as $\hat{T}=\mathrm{round}(N_x^{t}\tilde{\rho}_r)$, with a minimum length of 8 enforced to avoid degenerate cases. We use $\kappa=0.92$ for Mandarin targets and $\kappa=1$ otherwise to compensate for the systematic mismatch between acoustic and text-based syllable estimates.

\vspace{-0.5em}
\paragraph{Acoustic Syllable Estimation.}
To estimate $N_r^{a}$ without a reference transcript, we use prominent energy peaks within voiced regions as an acoustic proxy for syllables. Given a reference waveform $y$, we compute frame-level RMS energy using a 30-ms window and 10-ms hop, convert it to the dB scale as $e_m = 20\log_{10}(E_m+\epsilon)$, and use YIN~\citep{de2002yin} to identify voiced frames. We suppress the energy of unvoiced frames and retain peaks that exceed an adaptive energy threshold, have sufficient local prominence, and are separated from neighboring peaks by at least 100 ms. The number of retained peaks, $N_r^{a}=|\mathcal{P}|$, serves as the acoustic syllable-count estimate. The resulting pace is then clamped to $[7,12]$ to mitigate the effect of outliers.

\vspace{-0.5em}
\paragraph{Textual Syllable Estimation.}
We estimate the target syllable count $N_x^{t}$ directly from text. For Mandarin, each Chinese character is treated as one syllable. For English, we use the rule-based syllable estimator provided by \texttt{textstat}\footnote{textstat: \textcolor{magenta}{\url{https://github.com/textstat/textstat}}}. For mixed language text, the estimates from the two components are summed.

\subsubsection{Text-to-Semantic Distillation}
\label{sec:t2s-training}

To replace the AR T2S stage of IndexTTS2 with a substantially faster NAR architecture while preserving the downstream rendering pipeline, we train our T2S model through knowledge distillation from the original IndexTTS2 teacher rather than directly from raw speech corpora. Specifically, reference voices are paired with content-independent target text and passed through the teacher to synthesize the training data required by our model. This allows us to isolate the architectural change in T2S while retaining the teacher's semantic and acoustic representation spaces.

Training jointly optimizes two heads that share the same Transformer trunk: a masked-prediction head for semantic-code generation and a residual-recovery head that reconstructs the continuous representation expected by the downstream S2Mel renderer. The overall T2S objective is
\begin{equation}
\mathcal{L}_{\text{T2S}}
=
\mathcal{L}_{\text{mask}}
+
\lambda \mathcal{L}_{\text{res}}.
\end{equation}

\paragraph{Masked-prediction loss.}
For each training example, we sample a mask ratio $r=\cos(u)$, where $u\sim\mathcal{U}(0,\tfrac{\pi}{2})$. We then randomly replace $\lceil rT\rceil$ positions in the clean semantic-code sequence $\mathbf{y}$ with a \textsc{[mask]} token, yielding $\mathbf{y}^{(m)}$, and optimize cross-entropy only over the masked positions:
\begin{equation}
\mathcal{L}_{\text{mask}} = -\,\mathbb{E}\Big[
\textstyle\sum_{i:\,y_i^{(m)}=\textsc{mask}}
\log f_{\text{t2s}}^{\text{mask}}
\big(y_i \mid \mathbf{y}^{(m)}, E_{\text{txt}}, E_s, E_e \big)
\Big].
\end{equation}
Here, $f_{\text{t2s}}^{\text{mask}}$ denotes the masked-prediction head over semantic codes. We quantize the mask ratio into $16$ buckets and use the corresponding embedding as a timestep-like conditioning signal, which modulates every Transformer block through adaptive layer normalization (AdaLN)~\citep{peebles2023scalable}.

\paragraph{Residual-recovery loss.}
IndexTTS2 improves its S2Mel renderer by augmenting the discrete semantic-code embeddings with continuous GPT latent features extracted from the T2S model. We preserve this enhancement in our NAR T2S model by decomposing the teacher representation as
$S=\text{vq2emb}(\mathbf{y})+\Delta$,
where $\Delta$ denotes the additional continuous information beyond the discrete code embeddings. A residual head $f_{\text{t2s}}^{\text{res}}$, sharing the T2S Transformer backbone, is trained to recover $\Delta$ from the clean semantic sequence. Excluding the \textsc{eos}/\textsc{pad} tail, the residual-recovery loss is defined as:
\begin{equation}
\mathcal{L}_{\text{res}}
= \left\| f_{\text{t2s}}^{\text{res}}(\mathbf{y}, E_{\text{txt}}, E_s, E_e)
- \Delta \right\|_2^2,
\qquad
\Delta = S - \text{vq2emb}(\mathbf{y}).
\end{equation}

\subsection{Semantic-to-Audio Generation}
\label{sec:s2a}

The second stage converts the T2S output into waveform audio. We retain the S2Mel renderer architecture and conditioning interface of IndexTTS2, including its conditional flow-matching DiT and BigVGAN vocoder~\citep{lee2022bigvgan}. Our contribution in this stage is to reduce the renderer's inference cost: while the original model requires approximately $25$ Euler steps for high-quality synthesis, we reduce this to $4$--$8$ steps through reflow distillation.

\subsubsection{Semantic-to-Mel Reflow}
\label{sec:s2a-reflow}

We first fine-tune the pretrained IndexTTS2 renderer on the conditioning distribution produced by our T2S model, since its predicted semantic representation differs slightly from that of the original teacher. We then apply reflow~\citep{liu2022flow} to straighten the renderer's generation trajectories. Specifically, we run the fine-tuned renderer with its original multi-step solver and record the realized pairs between the initial Gaussian noise and the final generated mel spectrogram. The renderer is subsequently trained on these model-induced couplings, encouraging a straighter transport path that can be integrated accurately with substantially fewer Euler steps.

Both stages optimize the standard flow-matching velocity objective on the non-prompt mel region. For an interpolated state $y_t$ between noise $z$ and target mel $x_1$, the flow-matching DiT $g_{\text{s2mel}}$ predicts the velocity, and we minimize its mismatch with the target velocity $u_t$:
\begin{equation}
    \mathcal{L}_{\mathrm{flow}} = \mathbb{E} \left[
        \left\|
        g_{\text{s2mel}}(y_t,t,S,\cdot)_{[\ell,L)} - u_t{}_{[\ell,L)}
        \right\|_1
    \right].
\end{equation}
Here, $\ell$ and $L$ denote the prompt length and the total mel length in frames, so the loss covers only the non-prompt region $[\ell, L)$.
The fine-tuning stage uses fresh random noise and adapts the renderer to our T2S outputs, whereas the reflow stage uses the noise--output couplings recorded from the fine-tuned renderer itself. After reflow, the renderer achieves comparable synthesis quality using only $4$--$8$ Euler steps instead of approximately $25$.

\subsection{Training Data}
\label{sec:traindata}

We construct separate training sets for T2S and S2A distillation using IndexTTS2 as the teacher. For T2S, each sample pairs reference speech with a content-independent target sentence. A single teacher forward pass provides the semantic-code sequence $\mathbf{y}$, the latent features, the speaker, and emotion conditioning features. The dataset contains $874$k samples, about $2$k hours of speech from $2739$ unique reference speakers ($2342$ English speakers, $397$ Chinese), with target text decoupled from the reference content and approximately length-matched. Only the semantic token and latent features are stored since the loss function doesn't need the audio. For S2A, we build a separate coupling set using the fine-tuned renderer with its 25-step solver, conditioned on semantic representations from the trained T2S model. Each initial-noise/final-mel pair is stored as one coupling. The set contains $8$k pairs, corresponding to about $36$ hours of speech. Details are provided in Appendix~\ref{app:traindata}.

The training data is substantially smaller than the 55k hours used by IndexTTS2. Our goal is not to train a stronger model from scratch, but to improve inference efficiency while preserving as much of the teacher's performance as possible under a much smaller distillation budget. Despite this, \papername{} outperforms several baselines and achieves speaker similarity comparable to IndexTTS2 on English speech, which we attribute in part to our loss design that transfers both discrete semantic targets and continuous latent features from the teacher.
\section{Experiments} \label{sec:exp}

\subsection{Experiment Setup}
\label{sec:dataset}

\vspace{-0.3em}
\paragraph{Datasets.}
We evaluate on the four zero-shot test sets used by IndexTTS2~\citep{zhou2026indextts2}: LibriSpeech~\citep{panayotov2015librispeech} test-clean (English, $n{=}2544$), SeedTTS~\citep{anastassiou2024seed} test-en ($n{=}1088$) and test-zh ($n{=}2020$), and AISHELL-1~\citep{bu2017aishell} test (Mandarin, $n{=}1000$). We use the full sets without subsampling; each item pairs a reference clip with a content-independent target sentence, so the reference voice must be reproduced on unseen text. In addition, we demonstrate application scenarios of transcript-free models using three datasets: cross-lingual references in eight languages from FLEURS~\citep{conneau2023fleurs}, an infant-babble dataset of 70 audio recordings\footnote{Hugging Face Dataset: \textcolor{magenta}{\href{https://huggingface.co/datasets/Manisha12Researcher/Babies-weeping-and-happy-babbling-sounds}{\texttt{Babies\_weeping\_and\_happy\_babbling\_sounds}}}}, and a synthetic-gibberish set of 10 references. More details are in Appendices~\ref{app:multi-lingual} and~\ref{app:nonlex}.

\vspace{-0.3em}
\paragraph{Evaluation Metrics.}
Following the IndexTTS2 evaluation protocol, we measure \emph{intelligibility} using word error rate (WER) for English with Whisper~\citep{radford2023robust} and character error rate (CER) for Mandarin with FunASR~\citep{gao2023funasr}. We measure \emph{speaker similarity} (SS) as the cosine similarity between FunASR/CAM++~\citep{wang2023cam++} speaker embeddings of the generated and reference speech. For \emph{perceptual quality}, we use two no-reference MOS predictors: UTMOS~\citep{saeki2022utmos} and DNSMOS~\citep{reddy2022dnsmos}, we report its overall quality component. For \emph{efficiency}, we report the generation speed in $\times$ real-time, measured as seconds of generated audio per second of computation. All inference experiments are conducted on a single NVIDIA A100 GPU.

\vspace{-0.3em}
\paragraph{Baselines.}
We compare against the open zero-shot systems evaluated by IndexTTS2---CosyVoice2~\citep{du2024cosyvoice2scalablestreaming}, SparkTTS~\citep{wang2025spark}, MaskGCT~\citep{wang2025maskgct}, and F5-TTS~\citep{chen2025f5}---and against IndexTTS2 itself, our distillation teacher, shown together with its w/o-latent ablation as a reference upper bound rather than as a competitor. Quality numbers for all baselines on the four test sets are taken from the IndexTTS2 paper under identical metrics, while for the cross-lingual and non-lexical experiments we run all baselines ourselves, with each transcript-dependent baseline using the reference transcript produced by its own ASR front end; latency is measured by us in a common environment (SparkTTS omitted from timing, as it lacks a runnable release here). Throughout, \papername{} uses our default configuration: $12$ T2S unmasking steps, $8$ reflow S2Mel steps, and the transcript-free DSP length estimator (clamp $12$ codes/syllable).

\subsection{Main Results}
\begin{table*}[!h]
    \centering
    \fontsize{9}{11}\selectfont
    \setlength{\tabcolsep}{5pt}
    \resizebox{0.92 \textwidth}{!}{%
    \begin{tabular}{l cc cc cc cc}
    \hline
    & \multicolumn{2}{c}{LibriSpeech test-clean} & \multicolumn{2}{c}{SeedTTS test-en} & \multicolumn{2}{c}{SeedTTS test-zh} & \multicolumn{2}{c}{AISHELL-1 test} \\
    \cmidrule(lr){2-3}\cmidrule(lr){4-5}\cmidrule(lr){6-7}\cmidrule(lr){8-9}
    \textbf{Model} & SS$\uparrow$ & WER$\downarrow$ & SS$\uparrow$ & WER$\downarrow$ & SS$\uparrow$ & CER$\downarrow$ & SS$\uparrow$ & CER$\downarrow$ \\
    \hline
    {\textcolor{gray}{\textit{Autoregressive}}} & & & & & & & & \\
    CosyVoice2 & 0.843 & 5.999 & 0.794 & 3.277 & \textbf{0.846} & \textbf{1.451} & \textbf{0.834} & 1.967 \\
    SparkTTS   & 0.756 & 8.843 & 0.755 & \textbf{1.543} & 0.683 & 2.636 & 0.593 & \textbf{1.743} \\
    \hdashline
    {\textcolor{gray}{\textit{Non-autoregressive}}} & & & & & & & & \\
    MaskGCT & 0.790 & 7.759 & 0.824 & 2.530 & 0.807 & 2.447 & 0.598 & 4.930 \\
    F5-TTS  & 0.821 & 8.044 & 0.803 & 1.937 & 0.844 & 1.514 & 0.831 & 3.671 \\
    \textbf{\papername{}} & \textbf{0.875} & \textbf{5.54} & \textbf{0.849} & 2.34 & 0.829 & 1.71 & 0.804 & 2.21 \\
    \hdashline
    \multicolumn{2}{c}{\textcolor{gray}{\textit{Teacher (reference upper bound)}}} & & & & & & & \\
    \rowcolor{LLightGray} IndexTTS2 & 0.870 & 3.115 & 0.860 & 1.521 & 0.865 & 1.008 & 0.843 & 1.516 \\
    \rowcolor{LLightGray} IndexTTS2 w/o latent & 0.887 & 3.334 & 0.879 & 1.616 & 0.890 & 1.261 & 0.868 & 1.791 \\
    \hline
    \rowcolor{LightGray} \textit{Ground Truth} & 0.833 & 3.405 & 0.820 & 1.897 & 0.776 & 1.254 & 0.847 & 1.840 \\
    \hline
    \end{tabular}%
    }
    \caption{Zero-shot performance on the four datasets in speaker similarity (SS) and WER/CER. \textbf{Bold}: best per column among deployable systems (teacher and Ground Truth excluded). Results of all baselines are taken from the IndexTTS2 paper \citep{zhou2026indextts2}.}
    \label{tab:main}
    \end{table*}

We evaluate zero-shot voice cloning on four English and Mandarin test sets, covering speaker similarity and intelligibility. Table~\ref{tab:main} summarizes the results. Despite being distilled from IndexTTS2 and using a substantially simplified generation pipeline, \papername{} retains strong performance across both languages. The weaker Mandarin results may partly reflect the imbalance in speaker coverage ($2342$ English, $397$ Chinese). In particular, \papername{} achieves the best English speaker similarity among deployable systems and remains competitive with the strongest baselines on content accuracy. Overall, the results show that the distilled model preserves much of the teacher's zero-shot cloning capability while removing autoregressive decoding and reference-transcript dependence. We further evaluate perceptual quality using UTMOS and DNSMOS in Appendix~\ref{app:mos}, where \papername{} remains close to IndexTTS2 across all four test sets.

\subsection{Cloning from Cross-Lingual References} \label{sec:crossling}

We evaluate cross-lingual voice cloning, where the reference speech is in a language different from the target language and the reference transcript may be unreliable due to limited ASR support. We select speech samples in eight languages from the FLEURS dataset~\citep{conneau2023fleurs}: Japanese, Korean, German, Spanish, Russian, Arabic, Hindi, and Yoruba. For each language, we select three reference utterances from three different speakers. Each reference is used to generate speech for 10 English and 10 Chinese target sentences selected from the SeedTTS dataset. Table~\ref{tab:crossling-summary} reports the results averaged across the eight reference languages, while the per-language results are provided in Appendix~\ref{app:multi-lingual}. As shown in Table~\ref{tab:crossling-summary}, \papername{} achieves an SS of $0.789$ with $0.10\%$ WER on English targets, and an SS of $0.692$ with $1.90\%$ CER on Chinese targets, close to IndexTTS2. Its error rate remains below $3.4\%$ across all eight languages. In contrast, transcript-dependent baselines degrade substantially when the ASR front end does not support the reference language well, sometimes producing repeated or unrelated content or failing to generate speech. These results show that transcript-free conditioning enables more reliable cross-lingual voice cloning.

\begin{table}[!h]
    \centering
    \fontsize{9}{11}\selectfont
    \setlength{\tabcolsep}{5pt}
    \resizebox{0.88\textwidth}{!}{
    \begin{tabular}{l ccc ccc}
    \hline
    & \multicolumn{3}{c}{English targets} & \multicolumn{3}{c}{Chinese targets} \\
    \cmidrule(lr){2-4}\cmidrule(lr){5-7}
    \textbf{Model} & SS$\uparrow$ & WER$\downarrow$ & Fail$\downarrow$ & SS$\uparrow$ & CER$\downarrow$ & Fail$\downarrow$ \\
    \hline
    \multicolumn{3}{l}{\textcolor{gray}{\fontsize{7}{8}\textit{Transcript-dependent}}} & & & & \\
    F5-TTS     & 0.663{\scriptsize$\pm$0.120} & 24.56{\scriptsize$\pm$22.28} & 0.0{\scriptsize$\pm$0.0} & 0.622{\scriptsize$\pm$0.100} & 25.01{\scriptsize$\pm$20.89} & 0.0{\scriptsize$\pm$0.0} \\
    MaskGCT$^\dagger$ & 0.701{\scriptsize$\pm$0.065} & 20.16{\scriptsize$\pm$14.68} & 33.3{\scriptsize$\pm$47.1} & 0.700{\scriptsize$\pm$0.033} & 28.26{\scriptsize$\pm$14.32} & 33.3{\scriptsize$\pm$47.1} \\
    CosyVoice2 & 0.692{\scriptsize$\pm$0.081} & 45.95{\scriptsize$\pm$42.64} & 0.0{\scriptsize$\pm$0.0} & 0.680{\scriptsize$\pm$0.093} & 62.41{\scriptsize$\pm$47.33} & 0.0{\scriptsize$\pm$0.0} \\
    SparkTTS   & 0.611{\scriptsize$\pm$0.091} & 47.27{\scriptsize$\pm$42.23} & 20.0{\scriptsize$\pm$21.5} & 0.578{\scriptsize$\pm$0.092} & 60.42{\scriptsize$\pm$48.06} & 12.9{\scriptsize$\pm$16.9} \\
    \cdashline{1-7}
    \multicolumn{3}{l}{\textcolor{gray}{\fontsize{7}{8}\textit{Transcript-free}}} & & & & \\
    IndexTTS2  & \textbf{0.807}{\scriptsize$\pm$0.036} & 0.19{\scriptsize$\pm$0.28} & 0.0{\scriptsize$\pm$0.0} & \textbf{0.712}{\scriptsize$\pm$0.065} & \textbf{1.40}{\scriptsize$\pm$0.50} & 0.0{\scriptsize$\pm$0.0} \\
    \textbf{\papername{}} & 0.789{\scriptsize$\pm$0.041} & \textbf{0.10}{\scriptsize$\pm$0.19} & 0.0{\scriptsize$\pm$0.0} & 0.692{\scriptsize$\pm$0.073} & 1.90{\scriptsize$\pm$0.65} & 0.0{\scriptsize$\pm$0.0} \\
    \hline
    \end{tabular}
    }
    \caption{Results of 6 methods using cross-lingual references, reading English and Chinese targets. Results are averaged over eight reference languages, with $\pm$ indicating the standard deviation across languages. $\dagger$ MaskGCT fails on all Russian and Arabic references, so we report averages over the six languages with successful generation.}
    \label{tab:crossling-summary}
\end{table}

\subsection{Cloning from Non-Lexical References} \label{sec:nonlex}

\begin{table}[!h]
    \centering
    \fontsize{9}{11}\selectfont
    \setlength{\tabcolsep}{6pt}
    \resizebox{0.92 \textwidth}{!}{%
    \begin{tabular}{l cc c cc c cc c cc}
    \hline
    & \multicolumn{5}{c}{\textbf{Infant babble}} & & \multicolumn{5}{c}{\textbf{Synthetic gibberish}} \\
    \cline{2-6}\cline{8-12}
    & \multicolumn{2}{c}{English targets} & & \multicolumn{2}{c}{Chinese targets} & & \multicolumn{2}{c}{English targets} & & \multicolumn{2}{c}{Chinese targets} \\
    \cline{2-3}\cline{5-6}\cline{8-9}\cline{11-12}
    \textbf{Model} & SS$\uparrow$ & WER$\downarrow$ & & SS$\uparrow$ & CER$\downarrow$ & & SS$\uparrow$ & WER$\downarrow$ & & SS$\uparrow$ & CER$\downarrow$ \\
    \hline
    {\textcolor{gray}{\textit{Transcript-dependent}}} \\
    SparkTTS & 0.278 & 99.00 & & 0.310 & 96.00 & & 0.648 & 18.00 & & 0.639 & 20.30 \\
    CosyVoice2 & 0.442 & 31.03 & & 0.441 & 87.53 & & 0.721 & 49.24 & & 0.792 & 86.94 \\
    MaskGCT$^\dagger$ & 0.184 & 75.63 & & 0.217 & 45.67 & & 0.743 & 46.84 & & 0.729 & 32.81 \\
    F5-TTS & 0.315 & 43.36 & & 0.347 & 60.30 & & 0.699 & 17.33 & & 0.735 & 17.67 \\
    \hdashline
    {\textit{\textcolor{gray}{Transcript-free}}}  \\
    IndexTTS2 & 0.484 & 4.76 & & 0.425 & \textbf{1.34} & & \textbf{0.810} & \textbf{0.14} & & \textbf{0.880} & \textbf{1.23} \\
    \textbf{\papername{}} & \textbf{0.603} & \textbf{3.24} & & \textbf{0.595} & 5.79 & & 0.781 & 1.07 & & 0.846 & 1.82 \\
    \hline
    \end{tabular}%
    }
    \caption{Results of 6 methods using non-lexical references, reading English and Chinese targets. $^\dagger$MaskGCT results are averaged over the
    10 utterances that completed.}
    \label{tab:gibberish}
\end{table}

We further evaluate a more extreme setting where no meaningful reference transcript exists. We use infant babble and synthetic gibberish (e.g., ``kalene kuguva ge yu mu nola vi pesa'', see Table~\ref{tab:pseudowords}) as non-lexical references. Each reference is used to generate speech for 10 English and 10 Chinese target sentences selected from SeedTTS test-en and test-zh. Results are shown in Table~\ref{tab:gibberish}. Transcript-dependent baselines are affected by unreliable or empty ASR transcripts, sometimes resulting in unstable or failed generation. In contrast, both transcript-free systems generate speech for all references. \papername{} achieves higher speaker similarity than IndexTTS2 on infant babble for both English and Chinese targets, while both systems perform well on synthetic gibberish. These results show that transcript-free conditioning enables reliable voice cloning even when the reference contains no lexical content.

\subsection{Efficiency Analysis} \label{sec:eff}

Beyond wall-clock speed, the redesign also reduces the size of the generative model. \papername{} retains the teacher's conditioning encoder and BigVGAN vocoder, replaces the $527$M autoregressive T2S Transformer with a $269$M non-autoregressive masked generator, and further adapts the pretrained S2Mel renderer through fine-tuning and reflow. The T2S replacement therefore accounts for the model-size reduction and the majority of the latency improvement, while reflow provides additional acceleration in the S2A stage.

\begin{table}[!h]
    \centering
    \fontsize{9}{11}\selectfont
    \setlength{\tabcolsep}{6pt}
    \resizebox{0.78\textwidth}{!}{%
    \begin{tabular}{l rrrrr cc S[table-format=1.3]}
    \hline
    & \multicolumn{5}{c}{\textbf{Latency (ms)}} & \multicolumn{3}{c}{\textbf{Speed ($\times$ real-time)}$\uparrow$} \\
    \cmidrule(lr){2-6}\cmidrule(lr){7-9}
    \textbf{Model} & Cond. & ASR & \cellcolor{LightGray}\textbf{T2S} & \cellcolor{LightGray}\textbf{S2A} & Total & Cond. & \cellcolor{LightGray}\textbf{Gen.} & {Total} \\
    \hline
    {\textcolor{gray}{\textit{Transcript-dependent}}} \\
    CosyVoice2 & 1082 & 918 & \cellcolor{LightGray}3214 & \cellcolor{LightGray}604 & 5818 & 3.03 & \cellcolor{LightGray}1.56 & 1.03 \\
    MaskGCT    & 125 & 918 & \cellcolor{LightGray}2428 & \cellcolor{LightGray}2359 & 5830 & 5.88 & \cellcolor{LightGray}1.27 & 1.04 \\
    SparkTTS\,$^\ddagger$ & 57 & 918 & \cellcolor{LightGray}7041 & \cellcolor{LightGray}32 & 8049 & 6.67 & \cellcolor{LightGray}0.92 & 0.806 \\
    F5-TTS\,$^\dagger$ & 599 & 315 & \multicolumn{2}{>{\columncolor{LightGray}}c}{1122} & 2036 & 5.88 & \cellcolor{LightGray}4.76 & 2.63 \\
    \hdashline
    {\textcolor{gray}{\textit{Transcript-free}}} \\
    IndexTTS2 & 483 & - & \cellcolor{LightGray}5274 & \cellcolor{LightGray}875 & 6632 & 11.1 & \cellcolor{LightGray}0.88 & 0.813 \\
    \textbf{\papername{}} & 468 & - & \cellcolor{LightGray}{199} & \cellcolor{LightGray}310 & \textbf{977} & 11.1 & \cellcolor{LightGray}{9.09} & \bfseries \textbf{4.76~~} \\
    \hline
    \end{tabular}%
    }
    \caption{Per-stage latency (ms) and speed in seconds of audio generated per second of compute (s/s, the inverse of real-time factor) for condition computation, ASR, generation (T2S$+$S2A), and the overall pipeline. $^\dagger$F5-TTS is single-stage; its T2S/S2A are not separable. $^\ddagger$SparkTTS means are over the 31/32 items that generated audio---on one AISHELL item its decoder emitted no semantic tokens.}
    \label{tab:speed}
    \end{table}

We time every system on the same $32$ utterances ($8$ per test set, sampled to match each test set's mean length), on a single GPU with a global warm-up; outputs average about $5$\,s. Table~\ref{tab:speed} breaks per-utterance latency into three stages. Reference conditioning ($\sim$470\,ms) is a one-time cost that \papername{} inherits from the teacher unchanged, so the decisive difference lies in the T2S stage, where the autoregressive-to-non-autoregressive swap cuts latency from $5274$ to $199$\,ms, a $26.5\times$ reduction. Reflow distillation further trims S2A from 875 to 310 ms, so that generation, the sum of T2S and S2A, falls from 6149 to 509 ms, a 12.1× speedup that comes largely from the T2S replacement.

\begin{figure}[!h]
\centering
\includegraphics[width=0.9\linewidth]{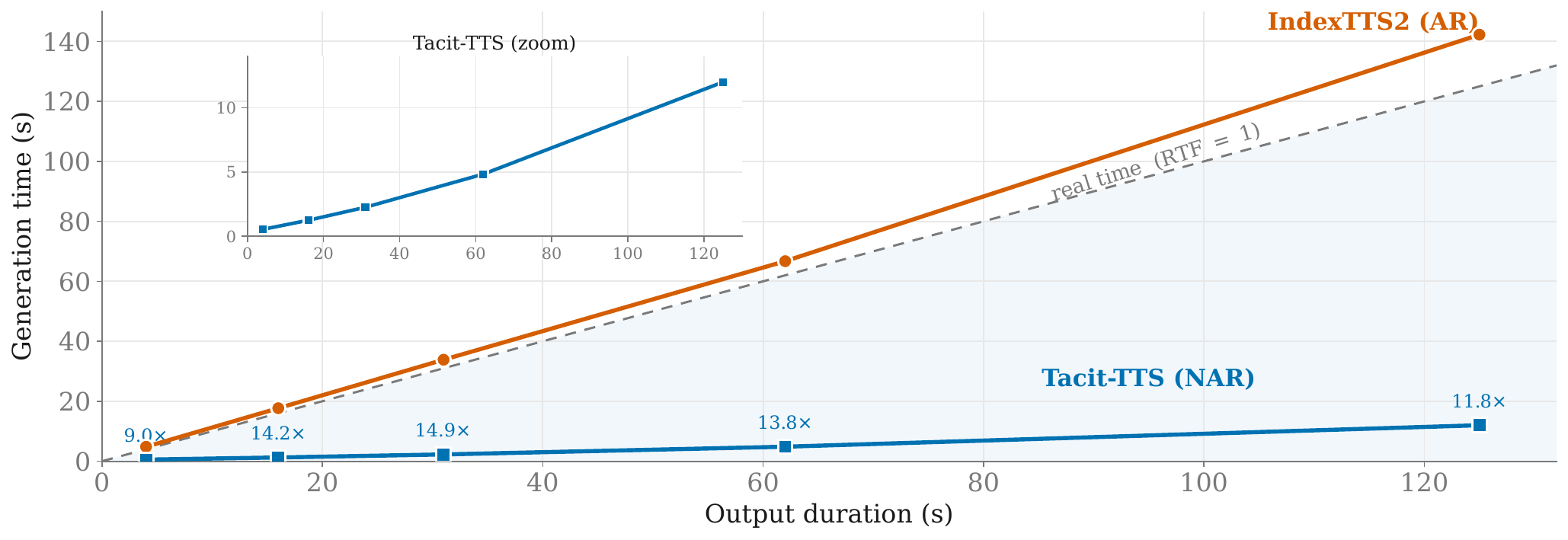}
\caption{End-to-end generation time (including reference conditioning) vs.\ output duration for IndexTTS2 and \papername{}, using one reference clip with a fixed passage repeated to lengthen the output. Dashed line: real time. Numbers above the \papername{} curve are its speedup over the teacher; the inset zooms in on \papername{}.}
\label{fig:length}
\vspace{-1em}
\end{figure}

\paragraph{Speedup vs.\ utterance length.}
The speedup of \papername{} over IndexTTS2 is not constant across generation duration (Figure~\ref{fig:length}). IndexTTS2's real-time factor is roughly flat ($\approx$1.1) because autoregressive cost is linear in output tokens. The speedup \emph{grows} from $9.0\times$ at four seconds to a peak of $14.9\times$ around thirty seconds, as the fixed per-utterance costs of both pipelines amortize over longer outputs. Past a minute it narrows again ($13.8\times$ at $62$\,s, $11.8\times$ at $125$\,s), as the full self-attention in the non-autoregressive stages ($O(n^2)$) begins to catch up. In practice, \papername{} generates roughly an order of magnitude faster than the teacher across the useful range, most of all at paragraph length, the regime where non-autoregressive generation helps most.

\paragraph{Generation Step Ablation}
We perform a grid search over the number of T2S unmasking steps and reflow S2Mel sampling steps, with T2S steps varied over ${8,12,16}$ and S2Mel steps over ${1,4,8}$; full results are provided in Table~\ref{tab:grid}. We select 12 T2S steps because increasing from 8 to 12 consistently improves content accuracy across datasets, while further increasing to 16 provides only marginal gains. For S2Mel, 8 steps are used because speaker similarity continues to improve from 4 to 8 steps, whereas fewer steps noticeably degrade speaker similarity. We therefore use the $12/8$ configuration as the default setting in all subsequent experiments.

\subsection{Duration Estimation} \label{sec:duration}
Our transcript-free length estimator determines the semantic length used for generation, so we evaluate whether it produces reasonable speech durations. Since natural speech inherently varies in duration even for the same utterance, we use Pearson correlation between generated and ground-truth durations rather than exact duration error. \papername{} achieves a correlation comparable to the other systems. Unlike learned duration predictors, our estimator is training-free and transcript-free, enabling generalization to out-of-distribution use cases (Sections~\ref{sec:crossling} and~\ref{sec:nonlex}). Detailed results are in Appendix~\ref{app:duration}.

\section{Conclusion}

We presented \papername{}, an efficient transcript-free zero-shot voice cloning system distilled from IndexTTS2. By combining masked non-autoregressive semantic generation, training-free length control, and an accelerated flow-matching renderer, \papername{} removes reference-transcript dependence while substantially improving inference efficiency. Across two English and two Mandarin datasets, it achieves competitive zero-shot voice cloning quality at a fraction of the teacher's generation cost. Beyond standard evaluation settings, \papername{} also remains effective with cross-lingual and non-lexical references, demonstrating the practical advantages of transcript-free conditioning. These results suggest that efficient zero-shot voice cloning can be achieved without relying on reference transcripts, opening new opportunities for voice cloning in scenarios where reliable transcripts are unavailable.

\section*{Ethics Statement}
Zero-shot voice cloning can be misused for impersonation, fraud, or the creation of non-consensual synthetic speech, and transcript-free cloning further lowers the barrier to cloning a voice from arbitrary recordings. Our work targets the efficiency of an existing publicly available voice cloning system and does not aim to extend what such systems can clone; nevertheless, we encourage any deployment of \papername{} to require consent from the cloned speaker and to be paired with safeguards such as audio watermarking and synthetic-speech detection. All speech used in this work comes from publicly released research datasets (LibriSpeech, AISHELL-1, ESD, the SeedTTS evaluation sets, FLEURS, and a public infant-vocalization collection), used under their respective licenses. We did not collect new recordings, and we make no attempt to identify any speaker.

\section*{Language Model Usage Statement}
In preparing this manuscript, we used a large language model only for grammar checking and minor language polishing. The authors reviewed and edited all suggestions. All scientific content, system design, experiments, analysis, and conclusions are entirely the work of the authors.

\appendix
\clearpage
\section*{\Large Appendix}
\counterwithin{figure}{section}
\counterwithin{equation}{section}
\counterwithin{table}{section}
\vspace{1em}

\section{Speaker and Emotion Conditioning Encoders} \label{app:percievers}
$E_s$ and $E_e$ are both extracted from the reference speech by IndexTTS2's own encoders, reused frozen, and both start from the same w2v-BERT-2.0 \citep{barrault2023seamless} features. $E_s$ is produced by a self-attention Conditioning Encoder followed by a Perceiver resampler \citep{jaegle2021perceiver} with $32$ learned latent queries, which cross-attends over the (projected) w2v-BERT-2.0 features to pool them into $32$ fixed-length latent tokens regardless of the reference clip's duration. $E_e$ is produced by a Conformer encoder followed by a single-latent Perceiver resampler (the same architecture, with only $1$ latent query instead of $32$) and two linear projection layers that map its native width back up to the shared conditioning width $d$.

\section{Non-Lexical Reference Set Construction} \label{app:nonlex}
\paragraph{Infant-babble references.}
The infant-babble dataset contains 70 audio recordings of infant babble from the collection footnoted in Section~\ref{sec:exp} (recordings of pre-linguistic infants vocalizing, $\sim$1\,min each, no accompanying text), from which we select 3 representative references as follows. Vocalization in these recordings occurs in short bursts, so we apply energy-based voice activity detection (RMS energy on $2048$-sample frames with a $512$-sample hop at 16\,kHz, a frame counting as voiced if its energy is within $30$\,dB of the audio's peak RMS frame) and keep only the longest continuous voiced region per recording; voiced region shorter than $3$\,s are dropped, leaving 24 segments. We embed each segment with the CAM++ speaker encoder, run $k$-means on the embeddings, choosing $k{=}3$ because it maximizes the silhouette coefficient ($0.55$ versus $\leq0.12$ for $k{=}4,5,6$), and select the segment closest to each cluster center.

\paragraph{Synthetic-gibberish references.}
We first build 10 pseudo-word sequences, shown in Table~\ref{tab:pseudowords}: each word is a string of $1$--$3$ syllables, each syllable formed by drawing one consonant and one vowel at random from 18 consonants and 5 vowels, with a fixed random seed, for $\sim$28 syllables per sequence. The sequences are pronounceable but contain no real word of any language, so no valid transcript exists. Each sequence is then fed as the target text to the teacher (IndexTTS2, 25 diffusion steps), conditioned on one of the 10 reference voice utterances in the official IndexTTS2 repo\footnote{\textcolor{magenta}{\url{https://github.com/index-tts/index-tts/tree/prepare_files/examples}}}. Table~\ref{tab:pseudowords} lists the 10 (prompt recording, pseudo-word sequence) pairs. 

\begin{table}[!h]
    \centering
    \fontsize{9}{10}\selectfont
    \setlength{\tabcolsep}{6pt}
    \resizebox{0.92 \textwidth}{!}{%
    \begin{tabular}{l p{9.5cm}}
    \hline
    \textbf{Prompt recording} & \textbf{Pseudo-word sequence (target text)} \\
    \hline
    voice\_01 & kalene kuguva ge yu mu nola vi pesa fa zula hukogu zuma ni na loze zela \\
    voice\_02 & tuneho nide so gese wohele lu fine wadape vugofu yiba likive biye kiyume \\
    voice\_03 & tuya sobazi de gawape wutiyu mumi zoyo ne sa nunaga ga guni wepuwe veka \\
    voice\_04 & vivoha kafike mu poti nahu du ge vo mode bolo vuwe re di du yuta geguge \\
    voice\_05 & kunu dugoyi mini pihi bo kamule gezi hu ruyara la pi mimi widavi bi le \\
    voice\_06 & huvuba pu zu popari ze nazuvu peteva so nitafa nehi nebe sigi yosa le \\
    voice\_07 & lakuvi soyafu la vayumi gisa ruroso rupevo turobi moso heyo tarusa \\
    voice\_08 & re pa no govumo fepa ve yo na poyu soyohe wolelu niha relisu pe fe mavosu \\
    voice\_09 & vamo bufo zi vuno lo boso topubo bavehe lo mosi lola budi ga ne pe wa molitu \\
    voice\_10 & kerabi fomana rukudi vigusa wavi heveyi wovuli gi nofi wi we lilu lubuma \\
    \hline
    \end{tabular}%
    }
    \caption{The 10 synthetic-gibberish text prompt.}
    \label{tab:pseudowords}
\end{table}

\section{Perceptual Quality Evaluation} \label{app:mos}
We report UTMOS and DNSMOS as reference measures of perceptual quality. Since IndexTTS2 is the teacher used for distillation, its scores serve as the main reference rather than a target to surpass. \papername{} remains close to the teacher across all four test sets, indicating that the distilled model largely preserves its perceptual quality. These automatic MOS predictors should be interpreted with caution. Some systems receive scores higher than the ground-truth recordings, suggesting that the predictors may favor speech distributions similar to their training data. We therefore use UTMOS and DNSMOS as supportive metrics rather than as definitive perceptual rankings.

\begin{table*}[!h]
    \centering
    \fontsize{9}{11}\selectfont
    \setlength{\tabcolsep}{5pt}
    \resizebox{0.98\textwidth}{!}{%
    \begin{tabular}{l cc cc cc cc}
    \hline
    & \multicolumn{2}{c}{LibriSpeech test-clean} & \multicolumn{2}{c}{SeedTTS test-en} & \multicolumn{2}{c}{SeedTTS test-zh} & \multicolumn{2}{c}{AISHELL-1 test} \\
    \cmidrule(lr){2-3}\cmidrule(lr){4-5}\cmidrule(lr){6-7}\cmidrule(lr){8-9}
    \textbf{Model} & {\fontsize{7}{8}\selectfont UTMOS$\uparrow$} & {\fontsize{7}{8}\selectfont DNSMOS$\uparrow$} & {\fontsize{7}{8}\selectfont UTMOS$\uparrow$} & {\fontsize{7}{8}\selectfont DNSMOS$\uparrow$} & {\fontsize{7}{8}\selectfont UTMOS$\uparrow$} & {\fontsize{7}{8}\selectfont DNSMOS$\uparrow$} & {\fontsize{7}{8}\selectfont UTMOS$\uparrow$} & {\fontsize{7}{8}\selectfont DNSMOS$\uparrow$} \\
    \hline
    {\textcolor{gray}{\textit{Autoregressive}}} & & & & & & & & \\
    CosyVoice2 & {4.344} & {3.348} & {4.115} & {3.253} & {3.395} & {3.382} & {2.980} & {3.251} \\
    SparkTTS & 4.017 & 3.065 & 3.903 & 3.133 & 3.271 & 3.266 & 2.864 & 3.081 \\
    \hdashline
    {\textcolor{gray}{\textit{Non-autoregressive}}} & & & & & & & & \\
    MaskGCT & 3.903 & 3.296 & 3.484 & 3.106 & 2.549 & 3.241 & 2.343 & 3.218 \\
    F5-TTS  & 3.102 & 3.304 & 3.592 & 3.235 & 2.864 & 3.353 & 2.069 & 3.162 \\
    \textbf{\papername{}} & 4.009 & 3.289 & 3.578 & 3.072 & 2.890 & 3.297 & 2.419 & 3.159 \\
    \hdashline
    \multicolumn{2}{c}{\textcolor{gray}{\textit{Teacher (reference upper bound)}}} & & & & & & & \\
    \rowcolor{LLightGray} IndexTTS2 & 4.051 & 3.319 & 3.572 & 3.077 & 2.926 & 3.303 & 2.486 & 3.285 \\
    \hline
    \rowcolor{LightGray} \textit{Ground Truth} & 4.087 & 3.275 & 3.457 & 3.073 & 2.690 & 3.214 & 2.337 & 3.207 \\
    \hline
    \end{tabular}%
    }
    \caption{Zero-shot performance on the four datasets in UTMOS and DNSMOS.}
    \label{tab:mos}
\end{table*}

\section{Generation Step Ablation}

\begin{table*}[!h]
    \centering
    \fontsize{9}{11}\selectfont
    \setlength{\tabcolsep}{5pt}
    \resizebox{0.92 \textwidth}{!}{%
    \begin{tabular}{cc cc cc @{\hspace{1.4em}} cc cc @{\hspace{1.2em}} c}
    \hline
    & & \multicolumn{4}{c}{{English}} & \multicolumn{4}{c}{{Chinese}} & \multirow{3}{*}{$1/\mathrm{RTF}$$\uparrow$} \\
    \cmidrule(lr){3-6}\cmidrule(lr){7-10}
    & & \multicolumn{2}{c}{LibriSpeech test-clean} & \multicolumn{2}{c}{SeedTTS test-en} & \multicolumn{2}{c}{SeedTTS test-zh} & \multicolumn{2}{c}{AISHELL-1 test} & \\
    \cmidrule(lr){3-4}\cmidrule(lr){5-6}\cmidrule(lr){7-8}\cmidrule(lr){9-10}
    {T2S} & {S2Mel} & SS$\uparrow$ & WER$\downarrow$ & SS$\uparrow$ & WER$\downarrow$ & SS$\uparrow$ & CER$\downarrow$ & SS$\uparrow$ & CER$\downarrow$ & \\
    \hline
    8  & 1 & 0.710 & 7.11 & 0.682 & 2.32 & 0.570 & 3.25 & 0.530 & 3.37 & 17.9 \\
    8  & 4 & 0.860 & 6.56 & 0.837 & 2.05 & 0.800 & 3.12 & 0.784 & 3.24 & 18.5 \\
    8  & 8 & 0.872 & 6.64 & 0.849 & 2.13 & 0.825 & 3.14 & 0.799 & 3.17 & 14.3 \\
    \hdashline
    12 & 1 & 0.711 & 5.78 & 0.684 & 2.36 & 0.575 & 1.81 & 0.535 & 2.39 & 19.2 \\
    12 & 4 & 0.863 & 5.63 & 0.838 & 2.25 & 0.804 & 1.69 & 0.789 & 2.26 & 15.6 \\
    \rowcolor{LLightGray} {12} & {8} & {0.875} & {5.54} & {0.849} & {2.34} & {0.829} & {1.71} & {0.804} & {2.21} & {12.5} \\
    \hdashline
    16 & 4 & 0.863 & 5.14 & 0.838 & 2.10 & 0.805 & 1.70 & 0.790 & 2.18 & 13.3 \\
    16 & 8 & 0.874 & 5.32 & 0.850 & 2.04 & 0.829 & 1.70 & 0.805 & 2.23 & 10.9 \\
    \hline
    \end{tabular}%
    }
    \caption{SS, WER/CER, and efficiency on LibriSpeech test-clean (mean duration: $7.1$,s) across T2S/S2Mel sampling steps. The gray row denotes the default configuration. Efficiency is $1/\mathrm{RTF}$ ($1\times$ = real time), excluding one-time conditioning computation.}
    \label{tab:grid}
    \end{table*}

\section{Training Data Construction} \label{app:traindata}
Each T2S training sample is generated by a single IndexTTS2 teacher forward pass, without synthesizing a waveform. We pair reference speech with a content-independent target sentence whose estimated syllable count is matched to the reference duration using corpus-specific speaking rates ($4.3$ syllables/s for English and $4.5$ for Mandarin). We store the teacher semantic codes, continuous latent $S$, speaker and emotion conditioning latents, and text tokens.

Reference speech is drawn from $2{,}334$ LibriSpeech~\citep{panayotov2015librispeech} training speakers, $389$ AISHELL-1~\citep{bu2017aishell} speakers, and $16$ ESD~\citep{zhou2021seen} speakers ($8$ English and $8$ Mandarin), excluding the four voices reserved for evaluation. Emotion coverage is further expanded using one-hot emotion conditioning. English target text consists of LibriSpeech training transcripts plus $6{,}000$ commentary-style sentences generated from hand-authored templates: sentence schemas over slot banks of players, teams, and match actions are composed at random, with exact-string deduplication ensuring no repeated sentence. Mandarin targets are assembled from AISHELL text fragments, randomly paired and joined into passages targeting a $2$--$15$\,s duration bucket at $4.5$ characters per second and capped at $70$ characters. Targeted supplements add (i) rare-word sentences: LibriSpeech transcript sentences containing words that failed in earlier evaluations, and Mandarin sentences built around such words; only samples the teacher pronounces correctly are kept, yielding $548$ English and $123$ Mandarin samples; (ii) $22{,}500$ ad-style commentary lines generated by combining hand-authored sentence templates with vocabularies of brand and athlete/team names, including known failure words ($1{,}940$ unique sentences); and (iii) $53{,}516$ extreme-length sentences built by inverting the main corpus length filter, spanning short content spans and long concatenations of LibriSpeech transcripts for English and of AISHELL fragments for Mandarin.

The final dataset contains $874{,}000$ samples, corresponding to about $1{,}960$ hours of semantic-code speech and $2{,}090$ hours per epoch after oversampling the rare-word and extreme-length subsets. It covers $2{,}739$ unique reference voices.

\paragraph{S2A coupling set.}
The reflow coupling set contains $8{,}000$ utterances generated in four parallel shards of $2{,}000$. Each utterance pairs a reference speaker, drawn from $400$ LibriSpeech and AISHELL-1 training speakers with an equal English/Chinese split, with a target sentence from the corresponding text pool. The trained T2S model produces the semantic representation, and the fine-tuned renderer integrates it from a recorded Gaussian noise sample with its 25-step Euler solver; each (initial noise, final mel) pair is stored as one coupling.

\section{Duration Fidelity Details} \label{app:duration}
\vspace{-1em}
\begin{figure}[!b]
\centering
\vspace{-2em}
\includegraphics[width=0.95\linewidth]{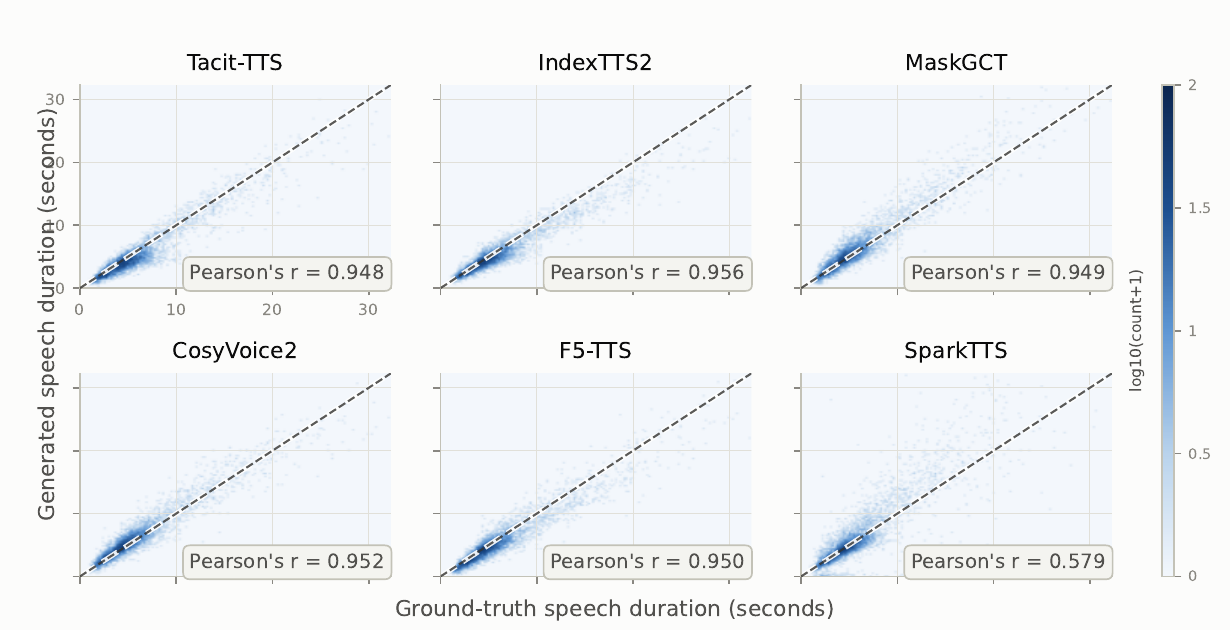}
\caption{Generated vs.\ ground-truth speech duration across the four zero-shot test sets. Each panel reports the pooled Pearson correlation $r$ for one model.}
\label{fig:duration}
\vspace{-1em}
\end{figure}

We compute Pearson's $r$ between the generated and ground-truth speech durations. The reported pooled result is computed over all utterances from the four test sets, jointly including both English and Mandarin samples. Figure~\ref{fig:duration} shows the corresponding scatter plots and per-test-set results.

\section{Cloning From Multi-lingual References} \label{app:multi-lingual}

\begin{table*}[!b]
\centering
\fontsize{9}{9}\selectfont
\setlength\arrayrulewidth{0.6pt}
\setlength{\tabcolsep}{4pt}
\resizebox{0.92 \textwidth}{!}{
\begin{tabular}{l cccccc cccccc}
\hline
 & \multicolumn{6}{c}{\textbf{Japanese}} & \multicolumn{6}{c}{\textbf{Korean}} \\
\cmidrule(lr){2-7}\cmidrule(lr){8-13}
 & \multicolumn{3}{c}{EN} & \multicolumn{3}{c}{ZH} & \multicolumn{3}{c}{EN} & \multicolumn{3}{c}{ZH} \\
\cmidrule(lr){2-4}\cmidrule(lr){5-7}\cmidrule(lr){8-10}\cmidrule(lr){11-13}
\textbf{Model} & SS$\uparrow$ & WER$\downarrow$ & Fail & SS$\uparrow$ & CER$\downarrow$ & Fail & SS$\uparrow$ & WER$\downarrow$ & Fail & SS$\uparrow$ & CER$\downarrow$ & Fail \\ \hline
\multicolumn{3}{l}{\textcolor{gray}{\fontsize{7}{8}\textit{Transcript-dependent}}} & & & & & & & & & & \\
F5-TTS & 0.408 & 38.86 & 0.0 & 0.406 & 40.99 & 0.0 & 0.648 & 31.33 & 0.0 & 0.643 & 23.59 & 0.0 \\
MaskGCT & 0.635 & 30.07 & 0.0 & 0.656 & 30.27 & 0.0 & 0.637 & 25.28 & 0.0 & 0.677 & 30.01 & 0.0 \\
CosyVoice2 & 0.629 & 25.04 & 0.0 & 0.603 & 78.62 & 0.0 & \textbf{0.746} & 3.81 & 0.0 & \textbf{0.819} & 8.21 & 0.0 \\
SparkTTS & 0.540 & 50.45 & 0.0 & 0.562 & 121.15 & 0.0 & 0.670 & 16.64 & 6.7 & 0.698 & 11.82 & 0.0 \\
\cdashline{1-13}
\multicolumn{3}{l}{\textcolor{gray}{\fontsize{7}{8}\textit{Transcript-free}}} & & & & & & & & & & \\
IndexTTS2 & \textbf{0.804} & 0.42 & 0.0 & \textbf{0.779} & \textbf{1.90} & 0.0 & 0.743 & \textbf{0.00} & 0.0 & 0.781 & \textbf{0.84} & 0.0 \\
\textbf{\papername{}} & 0.772 & \textbf{0.00} & 0.0 & 0.761 & 3.35 & 0.0 & 0.728 & \textbf{0.00} & 0.0 & 0.784 & 1.84 & 0.0 \\
\hline \hline
 & \multicolumn{6}{c}{\textbf{German}} & \multicolumn{6}{c}{\textbf{Spanish}} \\
\cmidrule(lr){2-7}\cmidrule(lr){8-13}
 & \multicolumn{3}{c}{EN} & \multicolumn{3}{c}{ZH} & \multicolumn{3}{c}{EN} & \multicolumn{3}{c}{ZH} \\
\cmidrule(lr){2-4}\cmidrule(lr){5-7}\cmidrule(lr){8-10}\cmidrule(lr){11-13}
\textbf{Model} & SS$\uparrow$ & WER$\downarrow$ & Fail & SS$\uparrow$ & CER$\downarrow$ & Fail & SS$\uparrow$ & WER$\downarrow$ & Fail & SS$\uparrow$ & CER$\downarrow$ & Fail \\ \hline
\multicolumn{3}{l}{\textcolor{gray}{\fontsize{7}{8}\textit{Transcript-dependent}}} & & & & & & & & & & \\
F5-TTS & 0.750 & 1.58 & 0.0 & 0.679 & 5.54 & 0.0 & 0.768 & 2.17 & 0.0 & 0.678 & 4.12 & 0.0 \\
MaskGCT & 0.791 & 7.94 & 0.0 & \textbf{0.747} & 48.53 & 0.0 & 0.764 & 15.35 & 0.0 & \textbf{0.713} & 27.06 & 0.0 \\
CosyVoice2 & 0.781 & 34.52 & 0.0 & 0.729 & 30.19 & 0.0 & 0.798 & 22.30 & 0.0 & 0.679 & 66.88 & 0.0 \\
SparkTTS & 0.686 & 12.11 & 33.3 & 0.588 & 3.07 & 33.3 & 0.729 & 21.38 & 0.0 & 0.677 & 51.77 & 0.0 \\
\cdashline{1-13}
\multicolumn{3}{l}{\textcolor{gray}{\fontsize{7}{8}\textit{Transcript-free}}} & & & & & & & & & & \\
IndexTTS2 & \textbf{0.800} & \textbf{0.00} & 0.0 & 0.689 & \textbf{1.22} & 0.0 & \textbf{0.844} & 0.67 & 0.0 & 0.689 & \textbf{0.96} & 0.0 \\
\textbf{\papername{}} & 0.784 & \textbf{0.00} & 0.0 & 0.678 & 2.09 & 0.0 & 0.830 & \textbf{0.42} & 0.0 & 0.692 & 1.27 & 0.0 \\
\hline \hline
 & \multicolumn{6}{c}{\textbf{Russian}} & \multicolumn{6}{c}{\textbf{Arabic}} \\
\cmidrule(lr){2-7}\cmidrule(lr){8-13}
 & \multicolumn{3}{c}{EN} & \multicolumn{3}{c}{ZH} & \multicolumn{3}{c}{EN} & \multicolumn{3}{c}{ZH} \\
\cmidrule(lr){2-4}\cmidrule(lr){5-7}\cmidrule(lr){8-10}\cmidrule(lr){11-13}
\textbf{Model} & SS$\uparrow$ & WER$\downarrow$ & Fail & SS$\uparrow$ & CER$\downarrow$ & Fail & SS$\uparrow$ & WER$\downarrow$ & Fail & SS$\uparrow$ & CER$\downarrow$ & Fail \\ \hline
\multicolumn{3}{l}{\textcolor{gray}{\fontsize{7}{8}\textit{Transcript-dependent}}} & & & & & & & & & & \\
F5-TTS & 0.653 & 57.31 & 0.0 & 0.632 & 64.80 & 0.0 & 0.767 & 14.99 & 0.0 & 0.622 & 11.33 & 0.0 \\
MaskGCT & - & - & 100.0 & - & - & 100.0 & - & - & 100.0 & - & - & 100.0 \\
CosyVoice2 & 0.609 & 67.05 & 0.0 & 0.581 & 89.64 & 0.0 & 0.581 & 82.40 & 0.0 & 0.568 & 66.67 & 0.0 \\
SparkTTS & 0.660 & 54.56 & 36.7 & 0.543 & 112.12 & 33.3 & 0.453 & 81.55 & 56.7 & 0.401 & 82.89 & 33.3 \\
\cdashline{1-13}
\multicolumn{3}{l}{\textcolor{gray}{\fontsize{7}{8}\textit{Transcript-free}}} & & & & & & & & & & \\
IndexTTS2 & 0.825 & \textbf{0.00} & 0.0 & \textbf{0.639} & 1.82 & 0.0 & \textbf{0.858} & \textbf{0.00} & 0.0 & \textbf{0.663} & \textbf{1.34} & 0.0 \\
\textbf{\papername{}} & \textbf{0.829} & \textbf{0.00} & 0.0 & 0.622 & \textbf{1.66} & 0.0 & 0.833 & 0.42 & 0.0 & 0.594 & 1.55 & 0.0 \\
\hline \hline
 & \multicolumn{6}{c}{\textbf{Hindi}} & \multicolumn{6}{c}{\textbf{Yoruba}} \\
\cmidrule(lr){2-7}\cmidrule(lr){8-13}
 & \multicolumn{3}{c}{EN} & \multicolumn{3}{c}{ZH} & \multicolumn{3}{c}{EN} & \multicolumn{3}{c}{ZH} \\
\cmidrule(lr){2-4}\cmidrule(lr){5-7}\cmidrule(lr){8-10}\cmidrule(lr){11-13}
\textbf{Model} & SS$\uparrow$ & WER$\downarrow$ & Fail & SS$\uparrow$ & CER$\downarrow$ & Fail & SS$\uparrow$ & WER$\downarrow$ & Fail & SS$\uparrow$ & CER$\downarrow$ & Fail \\ \hline
\multicolumn{3}{l}{\textcolor{gray}{\fontsize{7}{8}\textit{Transcript-dependent}}} & & & & & & & & & & \\
F5-TTS & 0.603 & 47.82 & 0.0 & 0.575 & 35.10 & 0.0 & 0.705 & 2.40 & 0.0 & 0.742 & 14.58 & 0.0 \\
MaskGCT & 0.698 & 40.87 & 66.7 & \textbf{0.688} & 30.00 & 66.7 & 0.679 & 1.42 & 0.0 & 0.720 & 3.72 & 0.0 \\
CosyVoice2 & 0.675 & 126.34 & 0.0 & 0.680 & 150.52 & 0.0 & 0.714 & 6.11 & 0.0 & 0.784 & 8.54 & 0.0 \\
SparkTTS & 0.567 & 131.13 & 0.0 & 0.542 & 91.32 & 0.0 & 0.580 & 10.31 & 26.7 & 0.611 & 9.23 & 3.3 \\
\cdashline{1-13}
\multicolumn{3}{l}{\textcolor{gray}{\fontsize{7}{8}\textit{Transcript-free}}} & & & & & & & & & & \\
IndexTTS2 & 0.798 & 0.48 & 0.0 & 0.656 & 2.17 & 0.0 & \textbf{0.785} & \textbf{0.00} & 0.0 & \textbf{0.803} & \textbf{0.95} & 0.0 \\
\textbf{\papername{}} & \textbf{0.798} & \textbf{0.00} & 0.0 & 0.633 & \textbf{2.04} & 0.0 & 0.740 & \textbf{0.00} & 0.0 & 0.772 & 1.38 & 0.0 \\
\hline
\end{tabular}
}
\caption{Per-language results of 6 methods using cross-lingual references, reading English (EN) and Chinese (ZH) targets. Fail: percentage of utterances with no generated speech. $\times$: no speech generated for that language. \textbf{Bold}: best SS and WER/CER per column.}
\label{tab:crosslingual}
\end{table*}

\paragraph{Reference set.}
References come from FLEURS~\citep{conneau2023fleurs}, a multilingual read-speech corpus recorded under a common protocol. None of the eight selected languages is a synthesis language of \papername{} or its teacher, both of which generate Chinese and English only. For each language we take three references of 6--12\,s, each from a different speaker: two of one gender and one of the other, alternating the majority gender across languages so that the set contains 12 male and 12 female speakers. Because FLEURS provides gender but not speaker identity, we follow the infant-babble procedure (Appendix~\ref{app:nonlex}): within each (language, gender) pool we run $k$-means on CAM++ speaker embeddings with $k$ equal to the number of references needed and take the clip closest to each centroid, further requiring a CAM++ cosine similarity below $0.6$ between any two references of the same language. References are taken from the FLEURS test split, except where it contains a single gender (German, Arabic, and Yoruba), in which case the other gender is taken from the training split.
\vspace{-1em}

\paragraph{Transcripts for transcript-dependent baselines.}
F5-TTS transcribes each reference with its built-in Whisper-large-v3-turbo model, while MaskGCT, CosyVoice2, and SparkTTS use Whisper-large-v3 transcripts with automatic language identification. MaskGCT additionally requires the prompt language; we pass the detected language when MaskGCT supports it (Japanese, Korean, and German here) and English otherwise, the only option available to a user of an unsupported language.


\begin{thebibliography}{43}
\providecommand{\natexlab}[1]{#1}
\providecommand{\url}[1]{\texttt{#1}}
\expandafter\ifx\csname urlstyle\endcsname\relax
  \providecommand{\doi}[1]{doi: #1}\else
  \providecommand{\doi}{doi: \begingroup \urlstyle{rm}\Url}\fi

\bibitem[Anastassiou et~al.(2024)Anastassiou, Chen, Chen, Chen, Chen, Chen,
  Cong, Deng, Ding, Gao, et~al.]{anastassiou2024seed}
Philip Anastassiou, Jiawei Chen, Jitong Chen, Yuanzhe Chen, Zhuo Chen, Ziyi
  Chen, Jian Cong, Lelai Deng, Chuang Ding, Lu~Gao, et~al.
\newblock Seed-tts: A family of high-quality versatile speech generation
  models.
\newblock \emph{arXiv preprint arXiv:2406.02430}, 2024.

\bibitem[Austin et~al.(2021)Austin, Johnson, Ho, Tarlow, and Van
  Den~Berg]{austin2021structured}
Jacob Austin, Daniel~D Johnson, Jonathan Ho, Daniel Tarlow, and Rianne Van
  Den~Berg.
\newblock Structured denoising diffusion models in discrete state-spaces.
\newblock \emph{Advances in neural information processing systems},
  34:\penalty0 17981--17993, 2021.

\bibitem[Barrault et~al.(2023)Barrault, Chung, Meglioli, Dale, Dong,
  Duppenthaler, Duquenne, Ellis, Elsahar, Haaheim,
  et~al.]{barrault2023seamless}
Lo{\"\i}c Barrault, Yu-An Chung, Mariano~Coria Meglioli, David Dale, Ning Dong,
  Mark Duppenthaler, Paul-Ambroise Duquenne, Brian Ellis, Hady Elsahar, Justin
  Haaheim, et~al.
\newblock Seamless: Multilingual expressive and streaming speech translation.
\newblock \emph{arXiv preprint arXiv:2312.05187}, 2023.

\bibitem[Borsos et~al.(2023)Borsos, Sharifi, Vincent, Kharitonov, Zeghidour,
  and Tagliasacchi]{borsos2023soundstorm}
Zal{\'a}n Borsos, Matt Sharifi, Damien Vincent, Eugene Kharitonov, Neil
  Zeghidour, and Marco Tagliasacchi.
\newblock Soundstorm: Efficient parallel audio generation.
\newblock \emph{arXiv preprint arXiv:2305.09636}, 2023.

\bibitem[Bu et~al.(2017)Bu, Du, Na, Wu, and Zheng]{bu2017aishell}
Hui Bu, Jiayu Du, Xingyu Na, Bengu Wu, and Hao Zheng.
\newblock Aishell-1: An open-source mandarin speech corpus and a speech
  recognition baseline.
\newblock In \emph{2017 20th conference of the oriental chapter of the
  international coordinating committee on speech databases and speech I/O
  systems and assessment (O-COCOSDA)}, pp.\  1--5. IEEE, 2017.

\bibitem[Casanova et~al.(2022)Casanova, Weber, Shulby, Junior, G{\"o}lge, and
  Ponti]{casanova2022yourtts}
Edresson Casanova, Julian Weber, Christopher~D Shulby, Arnaldo~Candido Junior,
  Eren G{\"o}lge, and Moacir~A Ponti.
\newblock Yourtts: Towards zero-shot multi-speaker tts and zero-shot voice
  conversion for everyone.
\newblock In \emph{International conference on machine learning}, pp.\
  2709--2720. PMLR, 2022.

\bibitem[Chang et~al.(2022)Chang, Zhang, Jiang, Liu, and
  Freeman]{chang2022maskgit}
Huiwen Chang, Han Zhang, Lu~Jiang, Ce~Liu, and William~T Freeman.
\newblock Maskgit: Masked generative image transformer.
\newblock In \emph{2022 IEEE/CVF Conference on Computer Vision and Pattern
  Recognition (CVPR)}, pp.\  11305--11315. IEEE, 2022.

\bibitem[Chen et~al.(2025)Chen, Niu, Ma, Deng, Wang, JianZhao, Yu, and
  Chen]{chen2025f5}
Yushen Chen, Zhikang Niu, Ziyang Ma, Keqi Deng, Chunhui Wang, JianZhao
  JianZhao, Kai Yu, and Xie Chen.
\newblock F5-tts: A fairytaler that fakes fluent and faithful speech with flow
  matching.
\newblock In \emph{Proceedings of the 63rd Annual Meeting of the Association
  for Computational Linguistics (Volume 1: Long Papers)}, pp.\  6255--6271,
  2025.

\bibitem[Chung et~al.(2021)Chung, Zhang, Han, Chiu, Qin, Pang, and
  Wu]{chung2021w2v}
Yu-An Chung, Yu~Zhang, Wei Han, Chung-Cheng Chiu, James Qin, Ruoming Pang, and
  Yonghui Wu.
\newblock W2v-bert: Combining contrastive learning and masked language modeling
  for self-supervised speech pre-training.
\newblock In \emph{2021 IEEE Automatic Speech Recognition and Understanding
  Workshop (ASRU)}, pp.\  244--250. IEEE, 2021.

\bibitem[Conneau et~al.(2023)Conneau, Ma, Khanuja, Zhang, Axelrod, Dalmia,
  Riesa, Rivera, and Bapna]{conneau2023fleurs}
Alexis Conneau, Min Ma, Simran Khanuja, Yu~Zhang, Vera Axelrod, Siddharth
  Dalmia, Jason Riesa, Clara Rivera, and Ankur Bapna.
\newblock Fleurs: Few-shot learning evaluation of universal representations of
  speech.
\newblock In \emph{2022 IEEE Spoken Language Technology Workshop (SLT)}, pp.\
  798--805. IEEE, 2023.

\bibitem[De~Cheveign{\'e} \& Kawahara(2002)De~Cheveign{\'e} and
  Kawahara]{de2002yin}
Alain De~Cheveign{\'e} and Hideki Kawahara.
\newblock Yin, a fundamental frequency estimator for speech and music.
\newblock \emph{The Journal of the Acoustical Society of America}, 111\penalty0
  (4):\penalty0 1917--1930, 2002.

\bibitem[De~Jong \& Wempe(2009)De~Jong and Wempe]{de2009praat}
Nivja~H De~Jong and Ton Wempe.
\newblock Praat script to detect syllable nuclei and measure speech rate
  automatically.
\newblock \emph{Behavior research methods}, 41\penalty0 (2):\penalty0 385--390,
  2009.

\bibitem[Deng et~al.(2025)Deng, Zhou, Shu, Wang, and Wang]{deng2025indextts}
Wei Deng, Siyi Zhou, Jingchen Shu, Jinchao Wang, and Lu~Wang.
\newblock Indextts: An industrial-level controllable and efficient zero-shot
  text-to-speech system.
\newblock \emph{arXiv preprint arXiv:2502.05512}, 2025.

\bibitem[Devlin et~al.(2019)Devlin, Chang, Lee, and Toutanova]{devlin2019bert}
Jacob Devlin, Ming-Wei Chang, Kenton Lee, and Kristina Toutanova.
\newblock Bert: Pre-training of deep bidirectional transformers for language
  understanding.
\newblock In \emph{Proceedings of the 2019 conference of the North American
  chapter of the association for computational linguistics: human language
  technologies, volume 1 (long and short papers)}, pp.\  4171--4186, 2019.

\bibitem[Du et~al.(2024{\natexlab{a}})Du, Chen, Zhang, Hu, Lu, Yang, Hu, Zheng,
  Gu, Ma, Gao, and Yan]{du2024cosyvoicescalablemultilingualzeroshot}
Zhihao Du, Qian Chen, Shiliang Zhang, Kai Hu, Heng Lu, Yexin Yang, Hangrui Hu,
  Siqi Zheng, Yue Gu, Ziyang Ma, Zhifu Gao, and Zhijie Yan.
\newblock Cosyvoice: A scalable multilingual zero-shot text-to-speech
  synthesizer based on supervised semantic tokens, 2024{\natexlab{a}}.
\newblock URL \url{https://arxiv.org/abs/2407.05407}.

\bibitem[Du et~al.(2024{\natexlab{b}})Du, Wang, Chen, Shi, Lv, Zhao, Gao, Yang,
  Gao, Wang, Yu, Liu, Sheng, Gu, Deng, Wang, Zhang, Yan, and
  Zhou]{du2024cosyvoice2scalablestreaming}
Zhihao Du, Yuxuan Wang, Qian Chen, Xian Shi, Xiang Lv, Tianyu Zhao, Zhifu Gao,
  Yexin Yang, Changfeng Gao, Hui Wang, Fan Yu, Huadai Liu, Zhengyan Sheng, Yue
  Gu, Chong Deng, Wen Wang, Shiliang Zhang, Zhijie Yan, and Jingren Zhou.
\newblock Cosyvoice 2: Scalable streaming speech synthesis with large language
  models, 2024{\natexlab{b}}.
\newblock URL \url{https://arxiv.org/abs/2412.10117}.

\bibitem[Eskimez et~al.(2024)Eskimez, Wang, Thakker, Li, Tsai, Xiao, Yang, Zhu,
  Tang, Tan, et~al.]{eskimez2024e2}
Sefik~Emre Eskimez, Xiaofei Wang, Manthan Thakker, Canrun Li, Chung-Hsien Tsai,
  Zhen Xiao, Hemin Yang, Zirun Zhu, Min Tang, Xu~Tan, et~al.
\newblock E2 tts: Embarrassingly easy fully non-autoregressive zero-shot tts.
\newblock In \emph{2024 IEEE spoken language technology workshop (SLT)}, pp.\
  682--689. IEEE, 2024.

\bibitem[Gao et~al.(2023)Gao, Li, Wang, Luo, Shi, Chen, Li, Zuo, Du, Xiao,
  et~al.]{gao2023funasr}
Zhifu Gao, Zerui Li, Jiaming Wang, Haoneng Luo, Xian Shi, Mengzhe Chen, Yabin
  Li, Lingyun Zuo, Zhihao Du, Zhangyu Xiao, et~al.
\newblock Funasr: A fundamental end-to-end speech recognition toolkit.
\newblock \emph{arXiv preprint arXiv:2305.11013}, 2023.

\bibitem[Hartman et~al.(2017)Hartman, Peluzzo, Shadani, Chellquist, Weprin,
  Hunt, Smith-Benjamin, and Altschuler]{hartman2017devising}
Kasondra Hartman, Amanda Peluzzo, Sharon Shadani, Ian Chellquist, Samuel
  Weprin, Halley Hunt, Sarah Smith-Benjamin, and Eric~L Altschuler.
\newblock Devising a method to study if wernicke’s aphasia patients are aware
  that they do not comprehend language or speak it understandably.
\newblock \emph{Journal of Undergraduate Neuroscience Education}, 16\penalty0
  (1):\penalty0 E5, 2017.

\bibitem[Hinton et~al.(2015)Hinton, Vinyals, and Dean]{hinton2015distilling}
Geoffrey Hinton, Oriol Vinyals, and Jeff Dean.
\newblock Distilling the knowledge in a neural network.
\newblock \emph{arXiv preprint arXiv:1503.02531}, 2015.

\bibitem[Jaegle et~al.(2021)Jaegle, Gimeno, Brock, Vinyals, Zisserman, and
  Carreira]{jaegle2021perceiver}
Andrew Jaegle, Felix Gimeno, Andy Brock, Oriol Vinyals, Andrew Zisserman, and
  Joao Carreira.
\newblock Perceiver: General perception with iterative attention.
\newblock In \emph{International conference on machine learning}, pp.\
  4651--4664. PMLR, 2021.

\bibitem[Jreige et~al.(2009)Jreige, Patel, and Bunnell]{jreige2009vocalid}
Camil Jreige, Rupal Patel, and H~Timothy Bunnell.
\newblock Vocalid: Personalizing text-to-speech synthesis for individuals with
  severe speech impairment.
\newblock In \emph{Proceedings of the 11th international ACM SIGACCESS
  conference on Computers and accessibility}, pp.\  259--260, 2009.

\bibitem[Ju et~al.(2024)Ju, Wang, Shen, Tan, Xin, Yang, Liu, Leng, Song, Tang,
  Wu, Qin, Li, Ye, Zhang, Bian, He, Li, and Zhao]{ju2024naturalspeech}
Zeqian Ju, Yuancheng Wang, Kai Shen, Xu~Tan, Detai Xin, Dongchao Yang, Yanqing
  Liu, Yichong Leng, Kaitao Song, Siliang Tang, Zhizheng Wu, Tao Qin,
  Xiang-Yang Li, Wei Ye, Shikun Zhang, Jiang Bian, Lei He, Jinyu Li, and Sheng
  Zhao.
\newblock {NaturalSpeech} 3: zero-shot speech synthesis with factorized codec
  and diffusion models.
\newblock In \emph{Proceedings of the 41st International Conference on Machine
  Learning}, ICML'24. JMLR.org, 2024.

\bibitem[Kim et~al.(2021)Kim, Kong, and Son]{kim2021conditional}
Jaehyeon Kim, Jungil Kong, and Juhee Son.
\newblock Conditional variational autoencoder with adversarial learning for
  end-to-end text-to-speech.
\newblock In \emph{International Conference on Machine Learning}, pp.\
  5530--5540. PMLR, 2021.

\bibitem[Le et~al.(2023)Le, Vyas, Shi, Karrer, Sari, Moritz, Williamson,
  Manohar, Adi, Mahadeokar, et~al.]{le2023voicebox}
Matthew Le, Apoorv Vyas, Bowen Shi, Brian Karrer, Leda Sari, Rashel Moritz,
  Mary Williamson, Vimal Manohar, Yossi Adi, Jay Mahadeokar, et~al.
\newblock Voicebox: Text-guided multilingual universal speech generation at
  scale.
\newblock \emph{Advances in neural information processing systems},
  36:\penalty0 14005--14034, 2023.

\bibitem[Lee et~al.(2022)Lee, Ping, Ginsburg, Catanzaro, and
  Yoon]{lee2022bigvgan}
Sang-gil Lee, Wei Ping, Boris Ginsburg, Bryan Catanzaro, and Sungroh Yoon.
\newblock Bigvgan: A universal neural vocoder with large-scale training.
\newblock \emph{arXiv preprint arXiv:2206.04658}, 2022.

\bibitem[Lipman et~al.(2022)Lipman, Chen, Ben-Hamu, Nickel, and
  Le]{lipman2022flow}
Yaron Lipman, Ricky~TQ Chen, Heli Ben-Hamu, Maximilian Nickel, and Matt Le.
\newblock Flow matching for generative modeling.
\newblock \emph{arXiv preprint arXiv:2210.02747}, 2022.

\bibitem[Liu et~al.(2022)Liu, Gong, and Liu]{liu2022flow}
Xingchao Liu, Chengyue Gong, and Qiang Liu.
\newblock Flow straight and fast: Learning to generate and transfer data with
  rectified flow.
\newblock \emph{arXiv preprint arXiv:2209.03003}, 2022.

\bibitem[Panayotov et~al.(2015)Panayotov, Chen, Povey, and
  Khudanpur]{panayotov2015librispeech}
Vassil Panayotov, Guoguo Chen, Daniel Povey, and Sanjeev Khudanpur.
\newblock Librispeech: an asr corpus based on public domain audio books.
\newblock In \emph{2015 IEEE international conference on acoustics, speech and
  signal processing (ICASSP)}, pp.\  5206--5210. IEEE, 2015.

\bibitem[Peebles \& Xie(2023)Peebles and Xie]{peebles2023scalable}
William Peebles and Saining Xie.
\newblock Scalable diffusion models with transformers.
\newblock In \emph{2023 IEEE/CVF International Conference on Computer Vision
  (ICCV)}, pp.\  4172--4182. IEEE, 2023.

\bibitem[Radford et~al.(2023)Radford, Kim, Xu, Brockman, McLeavey, and
  Sutskever]{radford2023robust}
Alec Radford, Jong~Wook Kim, Tao Xu, Greg Brockman, Christine McLeavey, and
  Ilya Sutskever.
\newblock Robust speech recognition via large-scale weak supervision.
\newblock In \emph{International conference on machine learning}, pp.\
  28492--28518. PMLR, 2023.

\bibitem[Reddy et~al.(2022)Reddy, Gopal, and Cutler]{reddy2022dnsmos}
Chandan~KA Reddy, Vishak Gopal, and Ross Cutler.
\newblock Dnsmos p. 835: A non-intrusive perceptual objective speech quality
  metric to evaluate noise suppressors.
\newblock In \emph{ICASSP 2022-2022 IEEE international conference on acoustics,
  speech and signal processing (ICASSP)}, pp.\  886--890. IEEE, 2022.

\bibitem[Ren et~al.(2019)Ren, Ruan, Tan, Qin, Zhao, Zhao, and
  Liu]{ren2019fastspeech}
Yi~Ren, Yangjun Ruan, Xu~Tan, Tao Qin, Sheng Zhao, Zhou Zhao, and Tie-Yan Liu.
\newblock Fastspeech: Fast, robust and controllable text to speech.
\newblock \emph{Advances in neural information processing systems}, 32, 2019.

\bibitem[Ren et~al.(2020)Ren, Hu, Tan, Qin, Zhao, Zhao, and
  Liu]{ren2020fastspeech}
Yi~Ren, Chenxu Hu, Xu~Tan, Tao Qin, Sheng Zhao, Zhou Zhao, and Tie-Yan Liu.
\newblock Fastspeech 2: Fast and high-quality end-to-end text to speech.
\newblock \emph{arXiv preprint arXiv:2006.04558}, 2020.

\bibitem[Saeki et~al.(2022)Saeki, Xin, Nakata, Koriyama, Takamichi, and
  Saruwatari]{saeki2022utmos}
Takaaki Saeki, Detai Xin, Wataru Nakata, Tomoki Koriyama, Shinnosuke Takamichi,
  and Hiroshi Saruwatari.
\newblock Utmos: Utokyo-sarulab system for voicemos challenge 2022.
\newblock \emph{arXiv preprint arXiv:2204.02152}, 2022.

\bibitem[Shen et~al.(2024)Shen, Ju, Tan, Liu, Leng, He, Qin, Bian,
  et~al.]{shen2024naturalspeech}
Kai Shen, Zeqian Ju, Xu~Tan, Eric Liu, Yichong Leng, Lei He, Tao Qin, Jiang
  Bian, et~al.
\newblock Naturalspeech 2: Latent diffusion models are natural and zero-shot
  speech and singing synthesizers.
\newblock In \emph{International conference on learning representations},
  volume 2024, pp.\  698--722, 2024.

\bibitem[Wang et~al.(2023{\natexlab{a}})Wang, Chen, Wu, Zhang, Zhou, Liu, Chen,
  Liu, Wang, Li, et~al.]{wang2023neural}
Chengyi Wang, Sanyuan Chen, Yu~Wu, Ziqiang Zhang, Long Zhou, Shujie Liu, Zhuo
  Chen, Yanqing Liu, Huaming Wang, Jinyu Li, et~al.
\newblock Neural codec language models are zero-shot text to speech
  synthesizers.
\newblock \emph{arXiv preprint arXiv:2301.02111}, 2023{\natexlab{a}}.

\bibitem[Wang et~al.(2023{\natexlab{b}})Wang, Zheng, Chen, Cheng, and
  Chen]{wang2023cam++}
Hui Wang, Siqi Zheng, Yafeng Chen, Luyao Cheng, and Qian Chen.
\newblock Cam++: A fast and efficient network for speaker verification using
  context-aware masking.
\newblock \emph{arXiv preprint arXiv:2303.00332}, 2023{\natexlab{b}}.

\bibitem[Wang et~al.(2025{\natexlab{a}})Wang, Jiang, Ma, Zhang, Liu, Li, Liang,
  Zheng, Wang, Feng, Bian, Ye, Cheng, Yuan, Zhao, Zhu, Pan, Xue, Zhu, Chen, Li,
  Chen, Xie, Guo, and Xue]{wang2025spark}
Xinsheng Wang, Mingqi Jiang, Ziyang Ma, Ziyu Zhang, Songxiang Liu, Linqin Li,
  Zheng Liang, Qixi Zheng, Rui Wang, Xiaoqin Feng, Weizhen Bian, Zhen Ye,
  Sitong Cheng, Ruibin Yuan, Zhixian Zhao, Xinfa Zhu, Jiahao Pan, Liumeng Xue,
  Pengcheng Zhu, Yunlin Chen, Zhifei Li, Xie Chen, Lei Xie, Yike Guo, and Wei
  Xue.
\newblock Spark-tts: An efficient llm-based text-to-speech model with
  single-stream decoupled speech tokens, 2025{\natexlab{a}}.
\newblock URL \url{https://arxiv.org/abs/2503.01710}.

\bibitem[Wang et~al.(2025{\natexlab{b}})Wang, Zhan, Liu, Zeng, Guo, Zheng,
  Zhang, Zhang, Zhang, and Wu]{wang2025maskgct}
Yuancheng Wang, Haoyue Zhan, Liwei Liu, Ruihong Zeng, Haotian Guo, Jiachen
  Zheng, Qiang Zhang, Xueyao Zhang, Shunsi Zhang, and Zhizheng Wu.
\newblock Maskgct: Zero-shot text-to-speech with masked generative codec
  transformer.
\newblock In \emph{International Conference on Learning Representations},
  volume 2025, pp.\  47127--47150, 2025{\natexlab{b}}.

\bibitem[Zhou et~al.(2021)Zhou, Sisman, Liu, and Li]{zhou2021seen}
Kun Zhou, Berrak Sisman, Rui Liu, and Haizhou Li.
\newblock Seen and unseen emotional style transfer for voice conversion with a
  new emotional speech dataset.
\newblock In \emph{ICASSP 2021-2021 IEEE International Conference on Acoustics,
  Speech and Signal Processing (ICASSP)}, pp.\  920--924. IEEE, 2021.

\bibitem[Zhou et~al.(2026{\natexlab{a}})Zhou, Zhang, Ng, Zhao, Wang, and
  Ma]{zhou2026emotional}
Kun Zhou, You Zhang, Dianwen Ng, Shengkui Zhao, Hao Wang, and Bin Ma.
\newblock Emotional dimension control in language model-based text-to-speech:
  Spanning a broad spectrum of human emotions.
\newblock In \emph{ICASSP 2026 - 2026 IEEE International Conference on
  Acoustics, Speech and Signal Processing (ICASSP)}, pp.\  17257--17261,
  2026{\natexlab{a}}.
\newblock \doi{10.1109/ICASSP55912.2026.11462862}.

\bibitem[Zhou et~al.(2026{\natexlab{b}})Zhou, Zhou, He, Zhou, Wang, Deng, and
  Shu]{zhou2026indextts2}
Siyi Zhou, Yiquan Zhou, Yi~He, Xun Zhou, Jinchao Wang, Wei Deng, and Jingchen
  Shu.
\newblock Indextts2: A breakthrough in emotionally expressive and
  duration-controlled auto-regressive zero-shot text-to-speech.
\newblock In \emph{Proceedings of the AAAI Conference on Artificial
  Intelligence}, volume~40, pp.\  35139--35148, 2026{\natexlab{b}}.

\end{thebibliography}
\end{document}